\documentclass[manuscript,screen,review=false]{acmart}

\usepackage{multirow}
\usepackage{colortbl}
\usepackage{graphicx}
\usepackage{caption}
\usepackage{subcaption}
\usepackage{enumitem}
\usepackage{extarrows}
\usepackage[normalem]{ulem}
\useunder{\uline}{\ul}{}
\usepackage{xcolor}
\usepackage{amsfonts} 
\usepackage{amsmath}
\usepackage{amsthm}
\usepackage{algorithm}
\usepackage{algpseudocode}

\newcommand{\ebf}{\textbf{e}}

\newcommand{\hbf}{\textbf{h}}

\newcommand{\sbf}{\textbf{s}}

\newcommand{\xbf}{\textbf{x}}

\newcommand{\zbf}{\textbf{z}}

\newcommand{\Abf}{\textbf{A}}

\newcommand{\Dbf}{\textbf{D}}

\newcommand{\Ibf}{\textbf{I}}

\newcommand{\Lbf}{\textbf{L}}

\newcommand{\Pbf}{\textbf{P}}

\newcommand{\Sbf}{\textbf{S}}

\newcommand{\Ubf}{\textbf{U}}
\newcommand{\Vbf}{\textbf{V}}
\newcommand{\Wbf}{\textbf{W}}

\AtBeginDocument{%
  }

\setcopyright{acmlicensed}
\copyrightyear{2018}
\acmYear{2018}
\acmDOI{XXXXXXX.XXXXXXX}

\acmConference[Conference acronym 'XX]{Make sure to enter the correct
  conference title from your rights confirmation emai}{June 03--05,
  2018}{Woodstock, NY}
\acmISBN{978-1-4503-XXXX-X/18/06}

\begin{document}

\title{Position-aware Graph Transformer for Recommendation}

\author{Jiajia Chen}
\email{jia2chan@mail.ustc.edu.cn}
\affiliation{
    \institution{Institute of Dataspace, Hefei Comprehensive National Science Center}
    \streetaddress{No. 230, East Beijing Road, Hefei}
    \postcode{230027}
    \country{China}
}
\author{Jiancan Wu}
\email{wujcan@gmail.com}
\affiliation{
    \institution{MoE Key Laboratory of Brain-inspired Intelligent Perception and Cognition, University of Science and Technology of China}
    \streetaddress{443 Huangshan Road, Hefei}
    \postcode{230027}
    \country{China}
}
\author{Jiawei Chen}
\email{sleepyhunt@zju.edu.cn}
\affiliation{
    \institution{School of Computer Science and Technology, Zhejiang University}
    \country{China}
}
\author{Chongming Gao}
\email{chongming.gao@gmail.com}
\affiliation{
    \institution{MoE Key Laboratory of Brain-inspired Intelligent Perception and Cognition, University of Science and Technology of China}
    \country{China}
}
\author{Yong Li}
\email{liyong07@tsinghua.edu.cn}
\affiliation{
    \institution{Department of Electronic Engineering, Tsinghua University}
    \country{China}
}
\author{Xiang Wang}
\authornote{Xiang Wang is also affiliated with Institute of Artificial Intelligence, Institute of Dataspace, Hefei Comprehensive National Science Center.}
\email{xiangwang1223@gmail.com}
\affiliation{
    \institution{MoE Key Laboratory of Brain-inspired Intelligent Perception and Cognition, University of Science and Technology of China}
    \country{China}
}


\renewcommand{\shortauthors}{Jiajia Chen et al.}

\begin{abstract}
Collaborative recommendation fundamentally involves learning high-quality user and item representations from interaction data. Recently, graph convolution networks (GCNs) have advanced the field by utilizing high-order connectivity patterns in interaction graphs, as evidenced by state-of-the-art methods like PinSage and LightGCN. However, one key limitation has not been well addressed in existing solutions: capturing long-range collaborative filtering signals, which are crucial for modeling user preference. In this work, we propose a new graph transformer (GT) framework --- \textit{Position-aware Graph Transformer for Recommendation} (PGTR), which combines the global modeling capability of Transformer blocks with the local neighborhood feature extraction of GCNs.
The key insight is to explicitly incorporate node position and structure information from the user-item interaction graph into GT architecture via several purpose-designed positional encodings. The long-range collaborative signals from the Transformer block are then combined linearly with the local neighborhood features from the GCN backbone to enhance node embeddings for final recommendations. Empirical studies demonstrate the effectiveness of the proposed PGTR method when implemented on various GCN-based backbones across four real-world datasets, and the robustness against interaction sparsity as well as noise. 
\end{abstract}




\maketitle

\section{Introduction}
\label{intro}

At the core of modern collaborative recommendation is learning high-quality representations of users and items from user-item interaction data.
Matrix factorization~\cite{SteffenRendle2009BPRBP,koren2009matrix} offered a first step towards representation learning for recommendation, projecting user and item IDs into an embedding space and performing predictions based on these vectors.
Subsequent work built on these foundations, augmenting the single user or item ID by incorporating additional information into the representations, such as interaction history~\cite{he2018nais,koren2008factorization} and rich side information~\cite{rendle2010factorization,rendle2011fast}.
More recently, approaches based on graph convolution networks (GCNs) have garnered significant attention in recommendation~\cite{he2020lightgcn,wang2019neural}.
It explicitly converts collaborative filtering signals into high-order connectivity patterns in the interaction graph, refining node representations through recursive neighborhood information propagation, thereby achieving state-of-the-art performance on recommendation tasks.

\begin{figure}[t]
    \centering
    \includegraphics[width=0.9\linewidth]{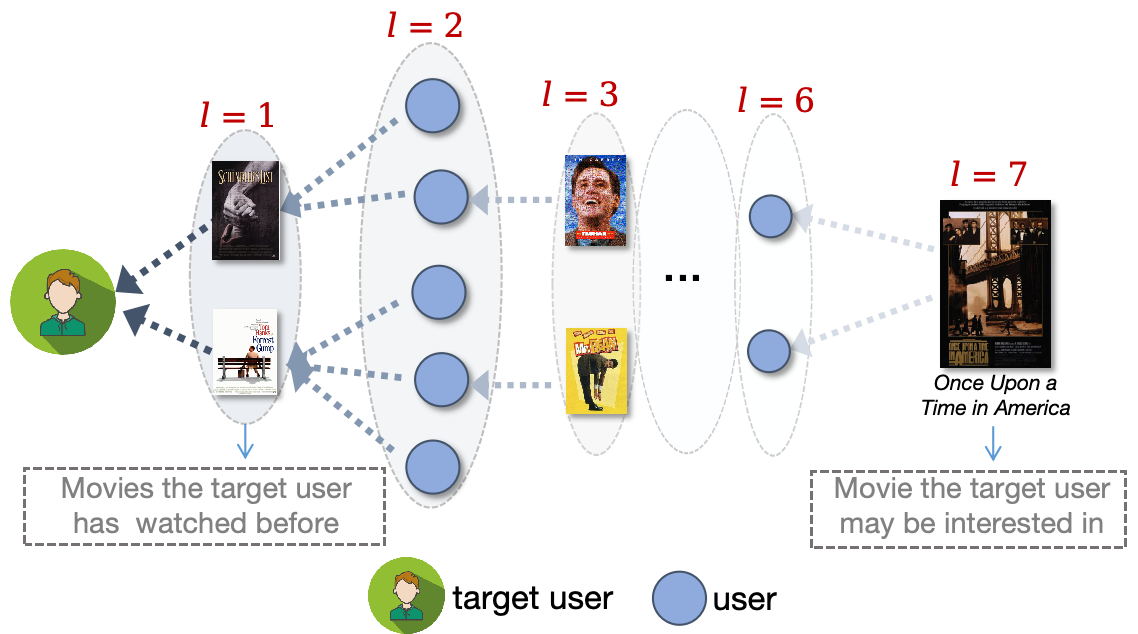}
    \caption{An example of long-range collaborative signals in movie recommendation, where the target user may be interested in the movie \textit{Once Upon a Time in America} seven hops away in the user-item graph. The dashed lines between nodes are the user-item interactions.}
    \label{fig:intro_longrange}
\end{figure}

Despite the success, we argue that existing GCN-based methods still suffer from a core shortcoming: \textbf{inadequate capability to capture long-range collaborative filtering signals}. The signals are pivotal in recommendation systems, as they implicitly contain a wealth of potential interactions that are useful for modeling user preferences. Figure \ref{fig:intro_longrange} shows an example of long-range collaborative signals in movie recommendation, where a less active target user might show interest in a movie seven hops away in the interaction graph. However, GCN-based methods, limited by well-recognized issues of over-smoothing~\cite{chen2020measuring} and over-squashing~\cite{alon2020bottleneck}, can only aggregate information from few hops (e.g., $l=2$ or 3) of neighbors~\cite{he2020lightgcn,wang2019neural,chen2020revisiting}. This limitation implies that many low-degree nodes may not be adequately modeled as they have difficulty capturing long-range collaborative signals. In addition, increasing the number of layers in GCNs might lead to the propagation of noise within the interaction graph, consequently having a detrimental effect on the user experience.


In this paper, we aim to answer the question "\textit{how to capture long-range collaborative filtering signals effectively and adequately?}". We focus on exploring of graph transformer (GT) technique in collaborative recommendation. The GT has been proven to effectively address the limitation of GCNs in capturing long-range information in the field of node/graph classification~\cite{ying2021transformers,rampavsek2022recipe,wu2022nodeformer,rong2020self}, but it is relatively less explored in recommendation. Typically, the GT framework combines GCNs with Transformer architectures, the former could encode local neighborhood features while the latter provides long-range/global information. At the core of the GT is positional encoding, which is crucial for capturing graph structure \cite{min2022transformer,muller2023attending}, so as to enhance the node representations. For example, Graphormer~\cite{ying2021transformers} devises positional encodings to capture the structure relation between nodes and the node importance in the graph. Compared with GCNs, the GT allows us to exploit the long-range/global information, achieving significant performance improvements in the classification tasks~\cite{min2022transformer,muller2023attending}.

To this end, we endeavor to introduce the advantages of the GT into recommendation. However, applying this technique to collaborative recommendation is not a trivial task, as the interaction graph data are extremely sparse, lack side information, and contain noise. Moreover, observed interactions often follow a power-law distribution~\cite{clauset2009power}, with significant variations in degree among different items or users. To address the aforementioned limitation of GCN-based recommenders, we propose a \textbf{P}osition-aware \textbf{G}raph \textbf{T}ransformer for \textbf{R}ecommendation (PGTR) on the basis of the current GCN model (e.g., LightGCN~\cite{he2020lightgcn}). Specifically, it is composed of two key components: \textbf{(1) well-designed positional encodings}, which effectively encode graph structure and node positional information on the user-item interaction graph, and \textbf{(2) global collaborative relations modeling}, which enables existing GCN-based models to adequately integrate long-range information for enhancing node representations. To be specific, the proposed positional encodings are tailored for the user-item graph in collaborative recommendation. They include \textit{spectral encoding}, \textit{degree encoding}, \textit{PageRank encoding}, and \textit{type encoding}. Among them, the \textit{spectral encoding} encodes the positional information of the nodes, the \textit{degree encoding} and \textit{PageRank encoding} consider the interaction characteristics of nodes on the graph, and the \textit{type encoding} focuses on the heterogeneity of the interaction graph. Each encoding works with a different rationality. This allows the individual information of nodes to be fully encoded on the interaction graph. After encoding position information for each node, the positional encodings are injected into the node representations. Subsequently, the Transformer block in PGTR models the relationships between all node pairs based on the position-aware node representations, effectuating the capture of global information. Finally, the captured global information is linearly combined with the local neighborhood features modeled by the GCN backbone, leading to the enhancement of the node representations.

It is worthwhile mentioning that our PGTR is model-agnostic. It
can be applied to any GCN-based model that consists of user and item embedding. In this work, we impletement it on NGCF~\cite{wang2019neural}, GCCF~\cite{chen2020revisiting}, LightGCN~\cite{he2020lightgcn} and UltraGCN~\cite{KelongMao2021UltraGCNUS}. Experimental studies on four benchmark datasets demonstrate the effectiveness of PGTR, which significantly improves the recommendation accuracy, and enhances the robustness against sparsity and noise. The contributions of this work are:

\begin{itemize}[leftmargin=*]
    \item We propose a model-agnostic GT framework, PGTR, on the basis of existing GCN-based models, which can capture long-range collaborative signals effectively and promote the performance of the GCN model.
    \item We highlight four node positional encodings tailored for the recommendation to provide interaction graph structure and node position information, which play a crucial role in our PGTR.
    \item  We conduct extensive experiments on multiple datasets to validate the effectiveness of our PGTR. By comparing it with several state-of-the-art methods, our approach consistently achieves improvements across various experimental settings.
\end{itemize}

\section{preliminaries}
In this section, we provide an introduction to basic symbols and fundamental methods, including the GCN-based collaborative filtering method and the GT framework.

Let $\mathbf{R} \in \mathbb{R}^{N\times M}$ denote the user-item interaction matrix constructed from dataset $\mathcal{D}$, where $N$ and $M$ represent the number of users and items, respectively. $\mathcal{U}$ and $\mathcal{I}$ represent the user set and item set respectively. Each entry $a_{ui}$ in $\mathbf{R}$ is a binary value, specifically 0 or 1, indicating whether user $u$ has interacted with item $i$. We can create a user-item bipartite graph $\mathcal{B} = (\mathcal{V}, \mathcal{E})$ based on the interaction matrix $\mathbf{R}$. Here, $\mathcal{V}$ consists of the set of user nodes and item nodes, while $\mathcal{E}$ represents the set of edges. An edge exists between user $u$ and item $i$ if $a_{ui} = 1$. $\mathcal{N}_u$ ($\mathcal{N}_i$) indicates the first-order neighbor items (users). $Degree(i)$ is the degree of node $i$ on the graph $\mathcal{B}$. $\mathbf{A}=\left(\begin{array}{cc}\mathbf{0} & \mathbf{R} \\ \mathbf{R}^\top & \mathbf{0}\end{array}\right)$ is the adjacency matrix.

\subsection{GCN-based Collaborative Filtering}
\label{sec_GCF}
The GCN-based method is a mainstream approach in recommendation~\cite{wang2019neural,he2020lightgcn,wang2019knowledge}, which aggregates neighborhood information, enabling us to capture high-order connectivity patterns. The common paradigm of GCN-based models in collaborative recommendation is as follows \cite{wu2021self}:
\begin{equation}
\label{eq23831}
\left\{
\begin{aligned}
\ebf_u^{(l+1)}&=f_{combine}(\ebf_u^{(l)}, f_{aggregate}(\{\ebf_i^{(l)}|i\in \mathcal{N}_u\})), \\
\ebf_i^{(l+1)}&=f_{combine}(\ebf_i^{(l)}, f_{aggregate}(\{\ebf_u^{(l)}|u\in \mathcal{N}_i\})).
\end{aligned}
\right.
\end{equation}
For a user node $u$, by aggregating the representations of its neighbors $\{\ebf_i^{(l)}|i\in \mathcal{N}_u\}$ and incorporating its own representation $\ebf_u^{(l)}$ at the $l$-th layer, the updated representation $\ebf_u^{(l+1)}$ is obtained. This holds true for each item node as well. After $L$ rounds of aggregation, the GCN model encodes the $L$-order neighbors. Subsequently, we obtain node representations for $L$ layers, resulting in the final node representation through a readout function:
\begin{equation}
\label{eq23891}
    \ebf = f_{readout}(\{\ebf^{(l)}|l=0,\cdots,L\}),
\end{equation}
where $\ebf^{(0)}\in \mathbb{R}^d$ is the initialized emebddings, $d$ is the embedding size. To be specific, there are a number of designs for $f_{combine}(\cdot)$, $f_{aggregate}(\cdot)$ and $f_{readout}(\cdot)$. More details can be referred to the survey \cite{wu2022graph}.

Traditional GCN-based methods can only capture collaborative signals through message passing and neighbor aggregation. However, when it comes to capturing long-range collaborative signals, such as $L > 3$, GCN-based models often face the issues of over-smoothing and over-squashing \cite{chen2020measuring,alon2020bottleneck}. This may result in a performance decline and the long-range collaborative signals can not be captured. The GT framework has the potential to address these issues. Next, we introduce the GT framework.

\subsection{Graph Transformer}
\label{sec_gt}
The GT techniques have been applied to various tasks of graph learning, such as node classification \cite{zhao2021gophormer,chen2022nagphormer,wu2022nodeformer}, graph classification \cite{nguyen2022universal,kim2022pure}. Typically, the GT framework combines GCN module with the Transformer module to leverage their respective strengths. The GCN module is utilized for learning the local structure of the graph, while the Transformer provides global information. By combining these two components in various configurations, the GT can effectively model both local and global relationships, resulting in improved performance for specific tasks \cite{min2022transformer}.

As mentioned before, the GCN model has been introduced. In this part, we focus on introducing the vanilla Transformer architecture. Assuming that the input of node $i$ to the $l$-th Transformer layer is denoted as $\zbf_i^{(l)}$, we compute attention scores within the Transformer across the full-graph (including $T$ nodes) to estimate the latent relationships between node $i$ and other nodes. For simplicity, there is
\begin{equation}
\label{eq23841}
\tilde{a}_{i j}^{(l)}=\frac{\exp \left(\left(\Wbf_Q^{(l)}\zbf_i^{(l)}\right)^\top\left(\Wbf_K^{(l)}\zbf_j^{(l)}\right)\right)}{\sum_{t=1}^T \exp \left(\left(\Wbf_Q^{(l)}\zbf_i^{(l)}\right)^{\top}\left(\Wbf_K^{(l)}\zbf_t^{(l)}\right)\right)},
\end{equation}
then perform global aggregation of node information:
\begin{equation}
\label{eq23842}
\zbf_i^{(l+1)}=\sum_{j=1}^T \tilde{a}_{i j}^{(l)} \cdot\left(\Wbf_V^{(l)}\zbf_j^{(l)}\right),
\end{equation}
where $\Wbf_Q^{(l)}$, $\Wbf_K^{(l)}$ and $\Wbf_V^{(l)}$ are learnable parameters at the $l$-th Transformer layer. The updating for all $T$ nodes in one layer using Eqs. \eqref{eq23841}-\eqref{eq23842} requires prohibitive $\mathcal{O}(T^2)$ complexity \cite{wu2022nodeformer}. When dealing with a large graph (with up to 2M nodes), computation becomes challenging. In order to accelerate the computation, Nodeformer \cite{wu2022nodeformer} proposes an efficient approximate algorithm named kernelized message passing method (a.k.a. Nodeformer Convolution) that reduces the computational complexity from $\mathcal{O}(T^2)$ to $\mathcal{O}(T)$. In brief, Eqs. \eqref{eq23841}-\eqref{eq23842} are finally transformed into 
\begin{equation}
\label{eq23881}
\left(\zbf_i^{(l+1)}\right)^\top \approx \frac{\phi\left(\Wbf_Q^{(l)}\zbf_i^{(l)}\right)^\top \sum_{j=1}^T \phi\left(\Wbf_K^{(l)}\zbf_j^{(l)} \right) \cdot \left(\Wbf_V^{(l)}\zbf_j^{(l)}\right)^{\top}}{\phi\left(\Wbf_Q^{(l)}\zbf_i^{(l)}\right) \cdot \sum_{t=1}^T \phi\left(\Wbf_K^{(l)}\zbf_t^{(l)}\right)^\top},
\end{equation}
where $\phi(\cdot): \mathbb{R}^{h} \rightarrow \mathbb{R}^{m}$ is a low-dimensional feature map with random transformation \cite{wu2022nodeformer}. For example,
\begin{equation}
    \phi(\mathbf{x})=\frac{\exp (\frac{-\|\mathbf{x}\|_2^2}{2})}{\sqrt{m}}\left[\exp \left(\mathbf{w}_1^\top\mathbf{x} \right), \cdots, \exp \left(\mathbf{w}_m^\top\mathbf{x}\right)\right],
\end{equation}
where $\xbf\in\mathbb{R}^{h}$ is the input, $\mathbf{w}_k\in\mathbb{R}^{h} \sim \mathcal{N}\left(0, \Ibf_h\right)$ ($\Ibf_h$ is the identity matrix) is an i.i.d. sampled random transformation \cite{choromanski2020rethinking}. This makes global node aggregation on large-scale graphs feasible.

\section{METHODOLOGY}
In this section, we present our proposed Position-aware Graph Transformer for Recommendation (PGTR) framework, which incorporates GCN-based backbone (e.g., LightGCN~\cite{he2020lightgcn}, NGCF~\cite{wang2019neural}), Transformer module (here we use the Nodeformer Convolution~\cite{wu2022nodeformer} introduced in Sec.~\ref{sec_gt}) and well-designed node positional encodings. The GCN-based backbone encodes local neighborhood features, the Transformer module is used for capturing global collaborative relations, as well as positional encodings are utilized to provide structure and position information of nodes on the interaction graph within the Transformer block. 

We start by introducing four types of node positional encodings tailored for collaborative recommendation. Then, we give the detailed implementations for our PGTR framework combined with the GCN-based backbone and the Transformer module.

\begin{figure*}
    \centering
    \includegraphics[width=1.0\textwidth]{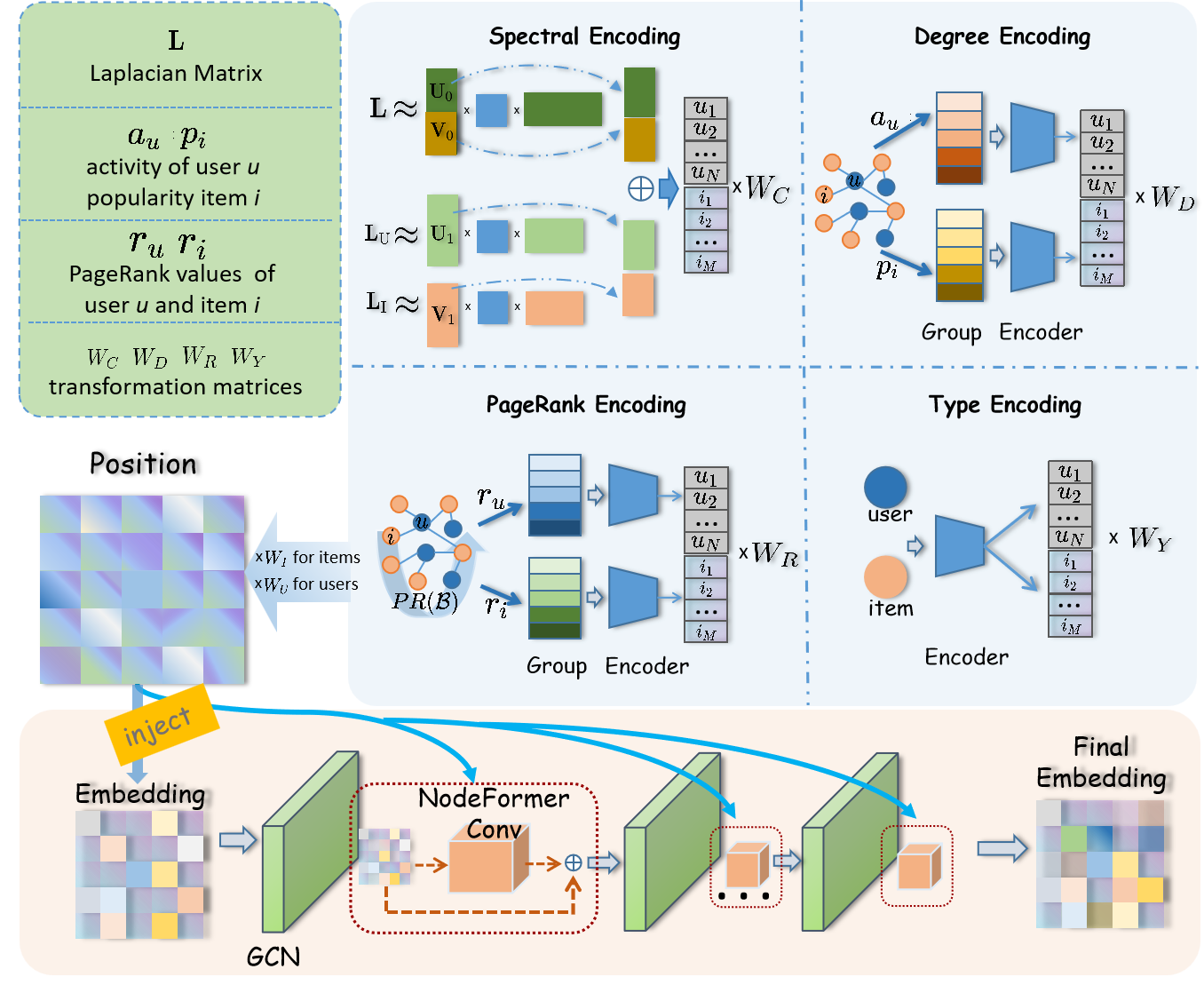}
    \caption{Overall framework of our PGTR model.}
    \label{fig:framework}
\end{figure*}

\subsection{Proposed Positional Encodings}

In practice, positional encodings are widely used in the GT for improving the capability of modeling the relations among nodes
\cite{cai2020graph, dwivedi2020generalization,zhang2020graph,kreuzer2021rethinking}. They provide the graph structure and position information of nodes within the Transformer module. However, they are not tailored for the field of collaborative recommendation. 
For example, it is crucial to not only consider the relationship between users and items but also take into account the relationship among users (user-user) and items (item-item). In addition, there are two types of nodes (user/item nodes) that need to be taken into consideration. In the next, we will introduce each of the positional encodings proposed for collaborative recommendation.

\subsubsection{\textbf{Spectral Encoding}} 
\label{sec_SPED}
In graph learning tasks, Laplacian positional encoding \cite{rampavsek2022recipe,park2022grpe,kreuzer2021rethinking,dwivedi2020generalization} is commonly utilized in the GT from the spectral domain of graph. This positional encoding can encode relative position information between nodes, and neighboring nodes would have similar encodings. In this paper, we adopt this method for encoding the position information of each user and item in the user-item interaction graph. Formally, for the interaction adjacency matrix $\Abf$, we precomputed its Laplacian eigenvectors,
\begin{equation}
\label{eq23861}
\mathbf{S}^\top \mathbf{\Lambda} \mathbf{S}=\Lbf=\mathbf{I}-\mathbf{D}^{-1 / 2} \mathbf{A} \mathbf{D}^{-1 / 2},
\end{equation}
where $\Dbf$ is the degree matrix of $\Abf$, $\Ibf$ is the identity matrix, $\mathbf{\Lambda}$ and $\Sbf$ are the eigenvalues and eigenvectors respectively. Generally, eigenvectors of the $H_C$ smallest non-trivial eigenvalues are used as the Laplacian positional encodings. Thus, we obtain the Laplacian positional encodings $\Pbf^{C_0}=\left[\mathbf{U}_0, \mathbf{V}_0
\right] \in \mathbb{R}^{H_C\times (N+M)}$ from $\Sbf$, where $\Ubf_0\in\mathbb{R}^{H_C\times N}$ and $\Vbf_0\in\mathbb{R}^{H_C\times M}$ represent the Laplacian positional encodings of users and items, respectively. 

In recommendation, building user-user and item-item relationships is of great significance~\cite{fan2023graph}. Therefore, we encode position information for each user and item on the user-side and item-side interaction graph, respectively. First, we construct the user-side adjacency matrix $\Abf_U \in \mathbb{R}^{N\times N}$ and the item-side adjacency matrix $\Abf_I \in \mathbb{R}^{M\times M}$, where the elements are 0 or 1, indicating the existence of second-order connectivity in $\mathcal{B}$. Then, we have 
\begin{equation}
\label{eq23871}
\left\{
\begin{aligned}
   &\mathbf{S}_1^\top \mathbf{\Lambda}_1 \mathbf{S}_1=\Lbf_U=\mathbf{I}_U-\mathbf{D}_U^{-1 / 2} \mathbf{A}_U \mathbf{D}_U^{-1 / 2},\\
  &\mathbf{S}_2^\top \mathbf{\Lambda}_2 \mathbf{S}_2=\Lbf_I=\mathbf{I}_I-\mathbf{D}_I^{-1 / 2} \mathbf{A}_I \mathbf{D}_I^{-1 / 2}.
\end{aligned}
\right.
\end{equation}
Likewise, we select the eigenvectors corresponding to the $H_C$ smallest non-trivial eigenvalues as the Laplacian positional encodings for users and items on their respective interaction graphs (user-side and item-side). That is, we obtain $\Ubf_1\in\mathbb{R}^{H_C\times N}$ from $\Sbf_1$, and $\Vbf_1\in\mathbb{R}^{H_C\times M}$ from $\Sbf_2$. 
Further, there is $\Pbf^{C_1}=\left[
\mathbf{U}_1,
\mathbf{V}_1
\right] \in \mathbb{R}^{H_C\times (N+M)}$. 

In the end, we combine $\Pbf^{C_0}$ and $\Pbf^{C_1}$ to create \textit{Spectral Encodings}, i.e.,
\begin{equation}
\label{eq24092202}
    \Pbf^C = (1-\lambda_C) \Pbf^{C_0} + \lambda_C \Pbf^{C_1},
\end{equation}
where $\lambda_C \in [0,1]$ is the hyper-parameter to control the proportions of different Laplacian positional encodings. Each user (item) can be determined by their spectral encodings separately on the user-item interaction graph and the user-side interaction graph (item-side interaction graph). This helps the model to have a clearer understanding of the relationships between users and items.

\subsubsection{\textbf{Degree Encoding}}
\label{sec_DSDE}
In recommendation systems, the popularity of items and the activity of users are important indicators that reflect the quality of items and user behavior patterns. From the perspective of items, when two items have similar popularity, they tend to appear in similar contexts or scenarios, such as in similar recommendation lists, purchase histories, or social circles. For example, some users are consistently drawn to popular songs, while others focus on niche or lesser-known tracks. Therefore, items with similar popularity are more likely to exhibit connections in user behavior data. From the perspective of users, when two users have similar activity, it implies that they exhibit similar behavior patterns and engagement on the platform. For example, based on similar preferences, they may collectively discover and explore similar items, or they may mutually follow each other on social networks. To this end, we devise degree encodings for each user and item.

\begin{figure*}
    \centering
    \includegraphics[width=1.0\textwidth]{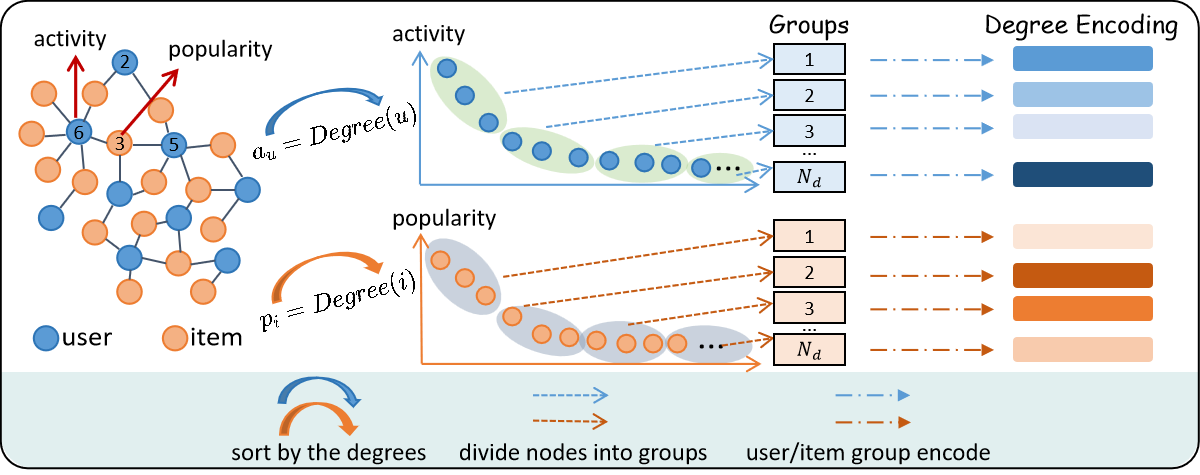}
    \caption{The procedure of Degree Encoding for users and items.}
    \label{fig:degree_encoding}
\end{figure*}

The degree distribution in recommendation datasets is skewed. For instance, in the Amazon-book dataset~\cite{he2020lightgcn}, some items have thousands of interactions, while others only have a dozen. If we encode each degree as done in Graphormer~\cite{ying2021transformers}, it becomes challenging to establish connections between nodes with close degrees. Therefore, we adopt a strategy of grouping degrees. 

Specifically, for the item set, each item $i \in \mathcal{I}$ has an associated popularity $p_i=Degree(i)$, i.e., 
\begin{equation}
\label{eq24092203}
    S_I = \{(i_1, p_{i_1}), (i_2, p_{i_2}), \cdots, (i_M, p_{i_M})\}.
\end{equation}
We sort the items in ascending order of their popularity and uniformly divide these items into $N_d$ groups. For an item $i$, the degree positional encoding is 
\begin{equation}
    \Pbf^D_i = DE_I(N_d^i), 0\leq N_d^i < N_d,
\end{equation}
where $DE_I(\cdot)$ is the degree group encoder for the items, $N_d^i$ is the group that the item $i$ belongs to. Finally, we obtain the degree encoding for all items, denoted as $\Pbf^D_I \in \mathbb{R}^{H_D\times M}$, $H_D$ is the encoding dimension.

Similarly, each user $u \in \mathcal{U}$ has a corresponding activity $a_u=Degree(u)$ and $S_U = \{(u_1, a_{u_1}), (u_2, a_{u_2}), \cdots, (u_N, a_{u_N})\}$. We sort the users in ascending order of their activity and also uniformly divide them into $N_d$ groups. The degree positional encoding of user $u$ is

\begin{equation}
\label{eq24092204}
    \Pbf^D_u = DE_U(N_d^u), 0\leq N_d^u < N_d,
\end{equation}
where $DE_U(\cdot)$ is the degree group encoder for the users, $N_d^u$ is the group that the user $u$ belongs to. Further, we can get $\Pbf^D_U \in \mathbb{R}^{H_D\times N}$ for all users.

It is necessary to note that this encoding is not entirely valid for items (or users) in the critical position within a group, because there are some items (or users) with similar popularity (activity) located in different groups. But in general, we make this positional encoding work by adjusting $N_d$.

\subsubsection{\textbf{PageRank Encoding}}
\label{sec_PGED}
PageRank \cite{page1998pagerank} can be used to describe the importance or influence of items and users on the interaction graph. The PageRank value of an item (or user) is not only related to its own degree but also depends on the importance of its neighbors on the graph. When two items have similar PageRank values, it indicates that they are similarly favored by users. For users, when two users have similar PageRank values, their interaction patterns tend to be similar. Therefore, we propose PageRank encodings.

\begin{figure*}
    \centering
    \includegraphics[width=1.0\textwidth]{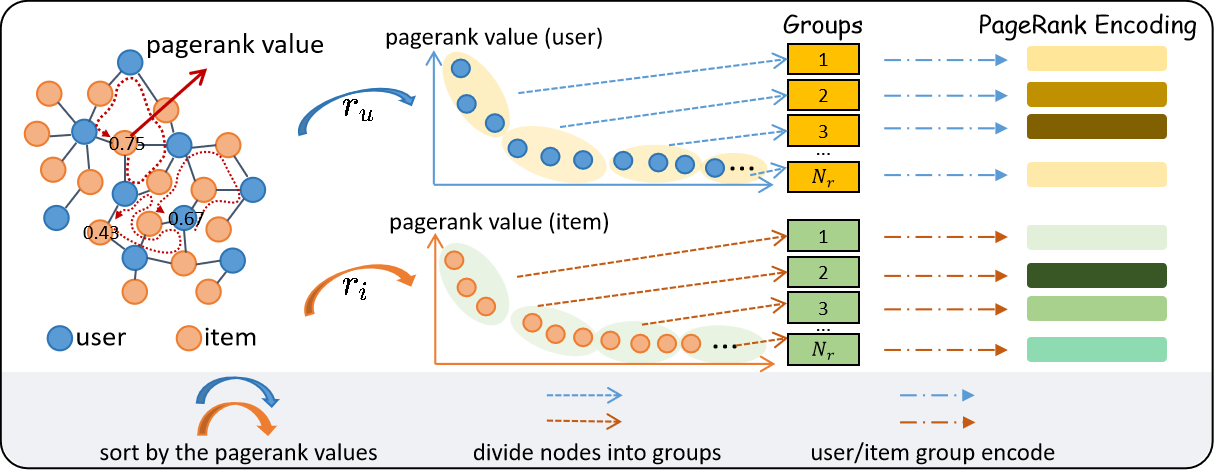}
    \caption{The procedure of PageRank Encoding for users and items.}
    \label{fig:pagerank_encoding}
\end{figure*}

First, we precompute PageRank values for all users and items based on the interaction graph $\mathcal{B}$ \footnote{We utilize \texttt{networkx.pagerank()} in Python to compute PageRank values.}. Then, each item $i$ has a PageRank value $r_i$ and there is 
\begin{equation}
\label{eq24092205}
    Q_I = \{(i_1, r_{i_1}), (i_2, r_{i_2}), \cdots, (i_M, r_{i_M})\}.
\end{equation}
We sort the items in ascending order of their PageRank values and divide these items into $N_r$ groups evenly. For an item $i$, its PageRank positional encoding is 
\begin{equation}
    \Pbf^R_i = PE_I(N_r^i), 0\leq N_r^i < N_r,
\end{equation}
where $PE_I(\cdot)$ is the PageRank group encoder for the items, $N_r^i$ is the group that the item $i$ belongs to. The PageRank positional encodings $\Pbf^R_I \in \mathbb{R}^{H_R\times M }$ for all items can be obtained, $H_R$ is the encoding dimension. Similar to the Section \ref{sec_DSDE}, from the user's perspective, we obtain the PageRank positional encodings corresponding to all users $\Pbf_{U}^R \in \mathbb{R}^{H_R\times N }$.

Implementing this encoding through grouping faces similar challenges as the \textit{Degree Similarity} criterion. However, by adjusting the number of groups, we also achieve its effectiveness in constructing the relationships among users and items. For simplicity, we set the number of groups for both items and users as $N_r$.

\subsubsection{\textbf{Type Encoding}}
\label{sec_TPED}
Users and items have different relation patterns and attributes in a recommendation system. For example, the display of items on a website or app (such as organized sorting by category or popularity) is different from the relationships between users. By distinguishing the node types on the interaction graph, the PGTR can better understand the interaction and connection among users and items. To this end, we propose type encodings based on the node types.

Intuitively, we assign a type of positional encoding to item groups and user groups separately. All items share the same type of positional encoding, and all users share the same one as well, i.e., 
\begin{equation}
\left\{
\begin{aligned}
    &\Pbf^Y_i = TE(0), i\in\mathcal{I},\\
    &\Pbf^Y_u = TE(1), u\in\mathcal{U}.
\end{aligned}  
\right.
\end{equation}
where $TE(\cdot)$ is the node type encoder. We can get $\Pbf^Y_I\in\mathbb{R}^{H_Y\times M}$ and $\Pbf^Y_U\in\mathbb{R}^{H_Y\times N}$ for all items and users respectively, $H_Y$ is the encoding dimension.

Overall, the proposed four positional encodings provide a thoughtful consideration of the relationship among users and items from different perspectives, enabling the capture and modeling of various relationships. Next, we introduce the PGTR framework and incorporate the four positional encodings into the framework to facilitate the learning of the relationships.

\subsection{PGTR Framework}
In this section, we integrate Eq. \eqref{eq23881} (i.e., Nodeformer Convolution) and several proposed positional encodings into the existing GCN-based model (e.g., LightGCN \cite{he2020lightgcn}, NGCF \cite{wang2019neural}) to form a new PGTR model. Figure \ref{fig:framework} illustrates the overall framework. Specifically, we first inject positional information into the initialization of node embeddings in the model, enabling position identification at the node representation. For  any given node $j$ (user or item), there is 
\begin{equation}\
\label{eq24092206}
   \hbf_j^{(0)} = \ebf_j^{(0)} + \lambda_1 \Pbf_j,
\end{equation}
where $\lambda_1\in [0,1]$ controls the strength of the positional information, and
\begin{equation}
\label{eq24092201}
\Pbf_j = \left\{
\begin{aligned}
     &\Wbf_I(\Wbf_C\Pbf^C_j + \Wbf_D\Pbf^D_j + \Wbf_R\Pbf^R_j + \Wbf_Y\Pbf^Y_j), \text{if}~j\in \mathcal{I}, \\
    &\Wbf_U(\Wbf_C\Pbf^C_j + \Wbf_D\Pbf^D_j + \Wbf_R\Pbf^R_j + \Wbf_Y\Pbf^Y_j), \text{if}~j\in \mathcal{U},
\end{aligned}
\right.
\end{equation}
where $\Wbf_I \in \mathbb{R}^{d\times d}$, $\Wbf_U \in \mathbb{R}^{d\times d}$, $\Wbf_C\in\mathbb{R}^{d\times H_C}$, $\Wbf_D\in\mathbb{R}^{d\times H_D}$, $\Wbf_R\in\mathbb{R}^{d\times H_R}$ and $\Wbf_Y\in\mathbb{R}^{d\times H_Y}$ are learnable transformation matrices that transform the positional encoding dimension to the embedding dimension. Next, $\hbf_j^{(0)}$ is fed to the GCN module and Nodeformer Convolution module. At the $l$-th Transformer layer, we have the following steps:
\begin{equation}
\label{eq24092207}
    \Tilde{\hbf}^{(l+1)}_j = f_{combine}(\hbf_j^{(l)}, f_{aggregate}(\{\hbf_k^{(l)}|k\in\mathcal{N}_j\})),
\end{equation}
this step models local neighborhood features through the GCN module. Thereafter, for capturing global collaborative signals, based on Eq. \eqref{eq23881},
\begin{equation}
\label{eq23882}
\left(\Bar{\hbf}_j^{(l+1)}\right)^\top \approx \frac{\phi\left(\Wbf_Q^{(l)}\hat{\hbf}_j^{(l)}\right)^\top\sum_{v=1}^{N+M} \phi\left(\Wbf_K^{(l)}\hat{\hbf}_v^{(l)}\right) \cdot \left(\Wbf_V^{(l)}\hat{\hbf}_v^{(l)}\right)^{\top}}{\phi\left(\Wbf_Q^{(l)} \hat{\hbf}_j^{(l)}\right)^{\top} \sum_{t=1}^{N+M} \phi\left(\Wbf_K^{(l)}\hat{\hbf}_t^{(l)}\right)},
\end{equation}
where
\begin{equation}
\label{eq24092208}
    \hat{\hbf}_j^{(l)} = \Tilde{\hbf}^{(l+1)}_j + \lambda_2 \Pbf_j,
\end{equation}
which means that the positional information is re-injected into the output representation of the GCN module, $\lambda_2\in [0,1]$ controls the injection strength as well. Finally, we get the representation of node $j$ at this layer,
\begin{equation}
\label{eq238101}
    \hbf^{(l+1)}_j = (1-\lambda_3)\Tilde{\hbf}^{(l+1)}_j + \lambda_3\Bar{\hbf}_j^{(l+1)},
\end{equation}
where $\lambda_3\in [0,1]$ controls the ratio of mixing between local and global information. After $L$ layers, there is the final node representation derived by our PGTR framework, i.e.,
\begin{equation}
\label{eq24092209}
    \hbf_j = f_{readout}(\{\hbf_j^{(l)}|l=0,\cdots,L\}).
\end{equation}  
Note that in our implementations, we omit $\Wbf_Q^{(l)}$, $\Wbf_K^{(l)}$ and $\Wbf_V^{(l)}$ in the Transformer for simplicity. 

$\bullet$ \textbf{Training Loss:} We train our PGTR model using the sampled softmax (SSM) loss \cite{wu2022effectiveness}, i.e.,
\begin{equation}
\label{eq24092210}
\mathcal{L}_{SSM}=-\frac{1}{|\mathcal{D}|} \sum_{(u, i) \in \mathcal{D}} \log \frac{\exp (f(u, i))}{\exp (f(u, i))+\sum_{j \in \mathcal{N}} \exp (f(u, j))},
\end{equation}
where $f(u, i)=\frac{\sbf_u^\top \sbf_i}{\tau}$, $\sbf_u$ and $\sbf_i$ are the normalized representations of $\hbf_u$ and $\hbf_i$ respectively, $\tau$ is the temperature coefficient, $j\in\mathcal{N}$ is the negative sample of user $u$. We adopt the in-batch negative trick \cite{DBLP:journals/corr/HidasiKBT15} just like \cite{wu2022effectiveness} for parallel computing, which treats positive items of other users in the same batch as the negatives. 

$\bullet$ \textbf{Space Complexity:} Note that among the four positional encodings, $\Pbf^D$, $\Pbf^R$ and $\Pbf^Y$ are trainable. Considering the space complexity of trainable parameters, the PGTR only adds a small number of trainable parameters compared to the original GCN-based model, i.e., $$\mathcal{O}\left(2(N_dH_D+N_rH_R+H_Y)+d(H_C+H_D+H_R+H_Y+2d)\right).$$
It is much smaller than all embedding parameters $\mathcal{O}((N+M)d)$ for large $N$ and $M$. 

Based on the observed experiments, for the two small datasets (Amazon-elec and Douban-book), the single-round training time of PGTR is similar to that of LightGCN. However, since PGTR establishes relationships between nodes extensively during training, it achieves faster performance improvement and requires fewer training rounds. On the larger LastFM and Amazon-book datasets, the single-round training time of PGTR is approximately 5x that of LightGCN. However, the overall training duration is only 1.2x longer on the LastFM dataset and 2.9x longer on the Amazon-book dataset. The comparison of computation efficiency between PGTR and the base model is shown in Section \ref{sec_compeff}. 

\begin{algorithm}
\caption{Training Algorithm of PGTR}
\begin{algorithmic}[1]
\State {\bfseries Data:} User-item interaction history $\mathcal{D}$ between $\mathcal{U}$ and $\mathcal{I}$; 
\State {\bfseries Input:} Batch size $B$, embedding size $d$, number of GCN layers $L$, Spectral encoding dimension $H_C$, Groups $N_d$, Degree encoding dimension $H_D$, Groups $N_r$, PageRank encoding dimension $H_R$, Type encoding dimension $H_Y$, \{$\lambda_C$, $\lambda_1$, $\lambda_2$, $\lambda_3$\};
\State {\bfseries Output:} Trained model parameters $\Theta=\{\mathbf{E}^{(0)}, \Wbf_Q^{(l)}, \Wbf_K^{(l)}, \Wbf_V^{(l)}, \Pbf^D, \Pbf^R, \Pbf^Y, \Wbf_I, \Wbf_U, \Wbf_C, \Wbf_D, \Wbf_R, \Wbf_Y\}$.
\State Initialize user and item embeddings $\mathbf{E}^{(0)}$(i.e., $\mathbf{e}^{(0)}_u, \mathbf{e}^{(0)}_i) \in \mathbb{R}^{d}$;
\State Construct user-item interaction matrix $\mathbf{R}$ from the dataset $\mathcal{D}$;
\State Compute adjacency matrix $\mathbf{A}$ and degree matrix $\mathbf{D}$ from $\mathbf{R}$;
\State Compute Spectral Encodings $\mathbf{P}^C$ (untrainable) for all nodes; \Comment{Section \ref{sec_SPED}} 
\State Compute Degree Encodings $\mathbf{P}^D$ (trainable) for all nodes; \Comment{Section \ref{sec_DSDE}}
\State Compute PageRank Encodings $\mathbf{P}^R$ (trainable) for all nodes; \Comment{Section \ref{sec_PGED}}
\State Compute Type Encodings $\mathbf{P}^Y$ (trainable) for all nodes; \Comment{Section \ref{sec_TPED}}
\State Initialize positional encodings $\mathbf{P}$ for all user and item nodes; \Comment{Eq. (\ref{eq24092201})}
\While{not done}
    \For{each batch $b \in \mathcal{D}$}
        \State Inject positional information $\mathbf{P}$ into the initialization of node embeddings $\mathbf{E}^{(0)}$; \Comment{Eq. (\ref{eq24092206})}
        \For{$l = 1$ to $L$}
            \State Model local neighborhood features through the GCN module; \Comment{Eq. (\ref{eq24092207})}
            \State Re-inject the positional information into the node representations; \Comment{Eq. (\ref{eq24092208})}
            \State Capture global collaborative signals through the Nodeformer Convolution module; \Comment{Eq. (\ref{eq23882})}
            \State Mix local and global information and get the representation of nodes at this layer; \Comment{Eq. (\ref{eq238101})}
        \EndFor
        \State Get the final node representations; \Comment{Eq. (\ref{eq24092209})}
        \State Calculate the sampled softmax loss $\mathcal{L}_{SSM}$ in the batch $b$; \Comment{Eq. (\ref{eq24092210})}
        \State Update model parameters $\Theta$ using gradient descent on the loss $\mathcal{L}_{SSM}$.
    \EndFor
\EndWhile
\end{algorithmic}
\end{algorithm}

\section{Experiments}

\renewcommand{\arraystretch}{1.5}
\begin{table*}[]
\centering
\caption{Performance comparison between our method PGTR and other counterparts. The best result is bolded and the runner-up is underlined. "[2020]" indicates the publication year of the corresponding method. Recall and NDCG scores are reported at a cutoff of 20.}
\begin{tabular}{c|cc|cc|cc|cc}
\hline
Data & \multicolumn{2}{c|}{Amazon-elec} & \multicolumn{2}{c|}{Douban-book} & \multicolumn{2}{c|}{LastFM} & \multicolumn{2}{c}{Amazon-book} \\ \hline
Model & Recall & NDCG & Recall & NDCG & Recall & NDCG & Recall & NDCG \\ \hline
MF {[}2009{]} & 0.0296 & 0.0410 & 0.0487 & 0.0837 & 0.0478 & 0.1054 & 0.0191 & 0.0361 \\ \hline
NGCF {[}2019{]} & 0.0684 & 0.0921 & 0.0735 & 0.1164 & 0.0263 & 0.0654 & 0.0141 & 0.0286 \\
GCCF {[}2020{]} & 0.0352 & 0.0534 & 0.0570 & 0.1114 & 0.0325 & 0.0840 & 0.0179 & 0.0364 \\
LightGCN {[}2020{]} & 0.0330 & 0.0485 & 0.0677 & 0.1278 & 0.0535 & 0.1215 & 0.0250 & 0.0472 \\
UltraGCN {[}2021{]} & {\ul 0.0781} & 0.0955 & {\ul 0.1123} & {\ul 0.1966} & {\ul 0.0612} & {\ul 0.1361} & 0.0212 & 0.0398 \\ \hline
EGLN {[}2021{]} & 0.0551 & 0.0732 & 0.0726 & 0.1353 & 0.0543 & 0.1227 & 0.0238 & 0.0454 \\
SGL {[}2021{]} & 0.0355 & 0.0465 & 0.0804 & 0.1490 & 0.0551 & 0.1235 & {\ul 0.0260} & 0.0463 \\
SimGCL {[}2022{]} & 0.0381 & 0.0510 & 0.0809 & 0.1517 & 0.0540 & 0.1221 & 0.0245 & 0.0432 \\
LightGCL {[}2023{]} & 0.0432 & 0.0610 & 0.0682 & 0.0904 & 0.0570 & 0.1217 & 0.0250 & 0.0428 \\
GFormer {[}2023{]} & 0.0761 & {\ul 0.1055} & 0.0781 & 0.1468 & 0.0577 & 0.1238 & 0.0258 & {\ul 0.0489} \\ \hline
\rowcolor[HTML]{EFEFEF} 
NGCF-PGTR & 0.0802 & 0.1084 & 0.0909 & 0.1558 & 0.0616 & 0.1394 & 0.0274 & 0.0506 \\
\rowcolor[HTML]{EFEFEF} 
GCCF-PGTR & 0.0796 & 0.1081 & 0.0889 & 0.1584 & 0.0579 & 0.1258 & 0.0264 & 0.0476 \\
\rowcolor[HTML]{EFEFEF} 
LightGCN-PGTR & \textbf{0.0811} & \textbf{0.1103} & 0.0899 & 0.1536 & 0.0610 & 0.1360 & \textbf{0.0290} & \textbf{0.0521} \\
\rowcolor[HTML]{EFEFEF} 
UltraGCN-PGTR & 0.0809 & 0.1087 & \textbf{0.1434} & \textbf{0.2287} & \textbf{0.0629} & \textbf{0.1395} & 0.0216 & 0.0404 \\ \hline
improve & 3.84\% & 4.55\% & 27.69\% & 16.33\% & 2.78\% & 2.50\% & 11.54\% & 6.54\% \\ \hline
\end{tabular}
\label{tab_PerfComp}
\end{table*}

In this section, we conduct experiments to evaluate the performance of our proposed PGTR. Our experiments are expected to answer the following research questions:

\begin{itemize}[leftmargin=*]
    \item \textbf{RQ1}: How effective is the proposed PGTR compared to other state-of-the-art (SOTA) models?
    \item \textbf{RQ2}: How does the proposed method perform on different levels of sparsity and noisy data? 
    \item \textbf{RQ3}: Are the four proposed positional encodings effective, and is capturing long-range collaborative signals useful?
\end{itemize}

\renewcommand{\arraystretch}{1.4}
\begin{table}[]
\caption{Dataset Description}
\centering
\begin{tabular}{c|c|c|c|c}
\hline
Dataset & \#Users & \#Items & \#Interactions & Density \\ \hline
Amazon-elec & 1,435 & 1,522 & 35,931 & 0.062\% \\ \hline
Douban-book & 13,024 & 22,347 & 792,062 & 0.272\% \\ \hline
LastFM & 23,566 & 48,123 & 3,034,763 & 0.268\% \\ \hline
Amazon-book & 52,643 & 91,599 & 2,984,108 & 0.062\% \\ \hline
\end{tabular}
\label{tab_dataset}
\end{table}

\subsection{Experimental Setup}
We perform experiments on four commonly used real-world datasets to evaluate our method: Amazon-electronics, Douban-book, LastFM, and Amazon-book. Table \ref{tab_dataset} provides an overview of the statistics for these datasets. In order to simulate different levels of sparsity, we divide the dataset into 20\%/40\%/60\%/80\% as the training set. Within the training set, 20\% is used as a validation set to tune hyper-parameters. The remaining 80\%/60\%/40\%/20\% is treated as the testing set. Two widely-used evaluation metrics are adopted, i.e., Recall@20 and NDCG@20.

\subsubsection{Compared Methods}
To validate the superiority of our method, we compare it with several advanced baselines used for collaborative filtering, including several well-known Matrix-Factorization based and GCN-based methods: \textbf{BPRMF} \cite{koren2009matrix} \textbf{NGCF} \cite{wang2019neural}, \textbf{GCCF} \cite{chen2020revisiting}, \textbf{LightGCN} \cite{he2020lightgcn} and \textbf{UltraGCN} \cite{KelongMao2021UltraGCNUS}, and several SOTA self-supervised learning based methods:
\begin{itemize}[leftmargin=*]
    \item \textbf{SGL} \cite{wu2021self}: It employs various random data augmentation techniques, such as edge dropping, node dropping, and random walks, to generate diverse views of interaction structures. These views are then used for contrastive learning.
    \item \textbf{EGLN} \cite{yang2021enhanced}: It incorporates two modules: a node embedding learning module and a graph structure learning module. They are designed for learning from each other and enhancing the representations.
    \item \textbf{SimGCL} \cite{yu2022graph}: It enhances LightGCN by incorporating a noise-based augmentation at the user/item representation level and contrastive learning. 
    \item \textbf{LightGCL} \cite{cai2022lightgcl}: It introduces a graph augmentation strategy based on singular value decomposition (SVD) to extract global collaborative patterns effectively. Contrastive learning is conducted between the reconstructed graph and the original graph.
    \item \textbf{GFormer} \cite{DBLP:conf/sigir/LiXRY0023}: This model enables the construction of rationale-aware general augmentation through masked graph autoencoding.
\end{itemize}

\subsubsection{Hyper-parameter Settings}
For a fair comparison, the GCN-based method that requires specifying the number of GCN layers is set to 2 layers. The embedding size for all methods is set to 32. All models use the Adam optimizer \cite{kingma2014adam}. Except for the Amazon-elec dataset, which has a batch size of 64, the batch sizes for the other three datasets are set to 2048. For our PGTR method, $H_C$ is searched from \{1,10,50,100,150,200,250,300\}, with a default value of 50. $H_D$, $H_R$, and $H_Y$ are searched from \{2,4,8,16,32,64\}, with a default value of 4. $N_d$ and $N_r$ are searched from \{1,5,10,15,20\}, with a default value of 10. For simplicity, $\lambda_C$, $\lambda_1$, and $\lambda_2$ are all searched in \{0,1\}. And the temperature $\tau$ of the SSM loss is searched in a coarse grain range of $\{0.02, 0.1, 0.2,\cdots,1.0,1.2\}$.

\subsection{Performance Comparison (\textbf{RQ1})}
In Table \ref{tab_PerfComp}, we present a performance comparison between our proposed method and other methods, where the dataset splitting is training:testing=2:8. Specifically, we apply the PGTR method to four existing GCN-based recommender models, enabling each model to capture long-range/global collaborative signals. Overall, our proposed method achieves improved performance over the four GCN-based models and attains the best performance on various datasets in such a sparse scene. Regarding Table \ref{tab_PerfComp}, the following observations can be made.
1) Overall, the GCN-based method outperforms the MF method, which is consistent with the current mainstream conclusion. This indicates the importance of utilizing neighbor information and capturing local collaborative relations.
2) Methods based on self-supervised learning demonstrate strong recommendation performance. However, we also observe that in cases where the training set is very sparse, some self-supervised learning methods perform similarly to the GCN-based models, such as SimGCL on the Amazon-book dataset, where its performance is lower than the base model LightGCN. This may be attributed to over-fitting, suggesting that relying solely on local information is insufficient for sparse data.
3) Our proposed method achieves significant performance improvements across various backbones. On one hand, by incorporating the position information of nodes into the model, our method can better capture relations among nodes. On the other hand, employing the Transformer architecture allows the model to capture long-range/global collaborative signals, compensating for the limitations of the existing GCN-based models.

\subsection{Robustness Study (\textbf{RQ2})}
In this section, we evaluate the robustness of our model compared to other methods against data sparsity and noise. 
\subsubsection{Robustness Against Data Sparsity} 
\begin{figure}  
\centering
\begin{subfigure}[b]{0.45\textwidth}
    \centering    
    \includegraphics[width=\textwidth]{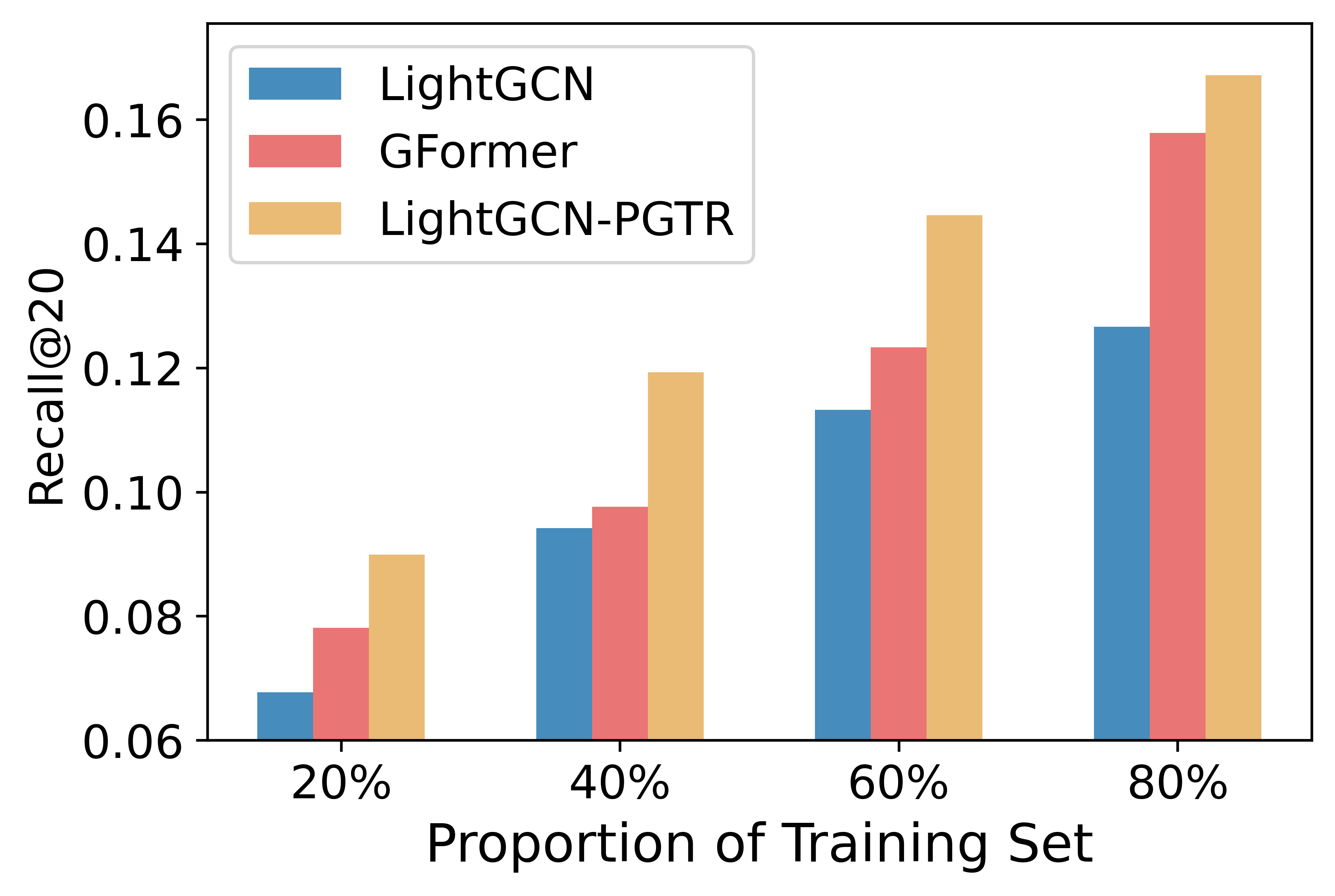}
    (a) Douban-book Recall
\end{subfigure}%
\begin{subfigure}[b]{0.45\textwidth}
    \centering
    \includegraphics[width=\textwidth]{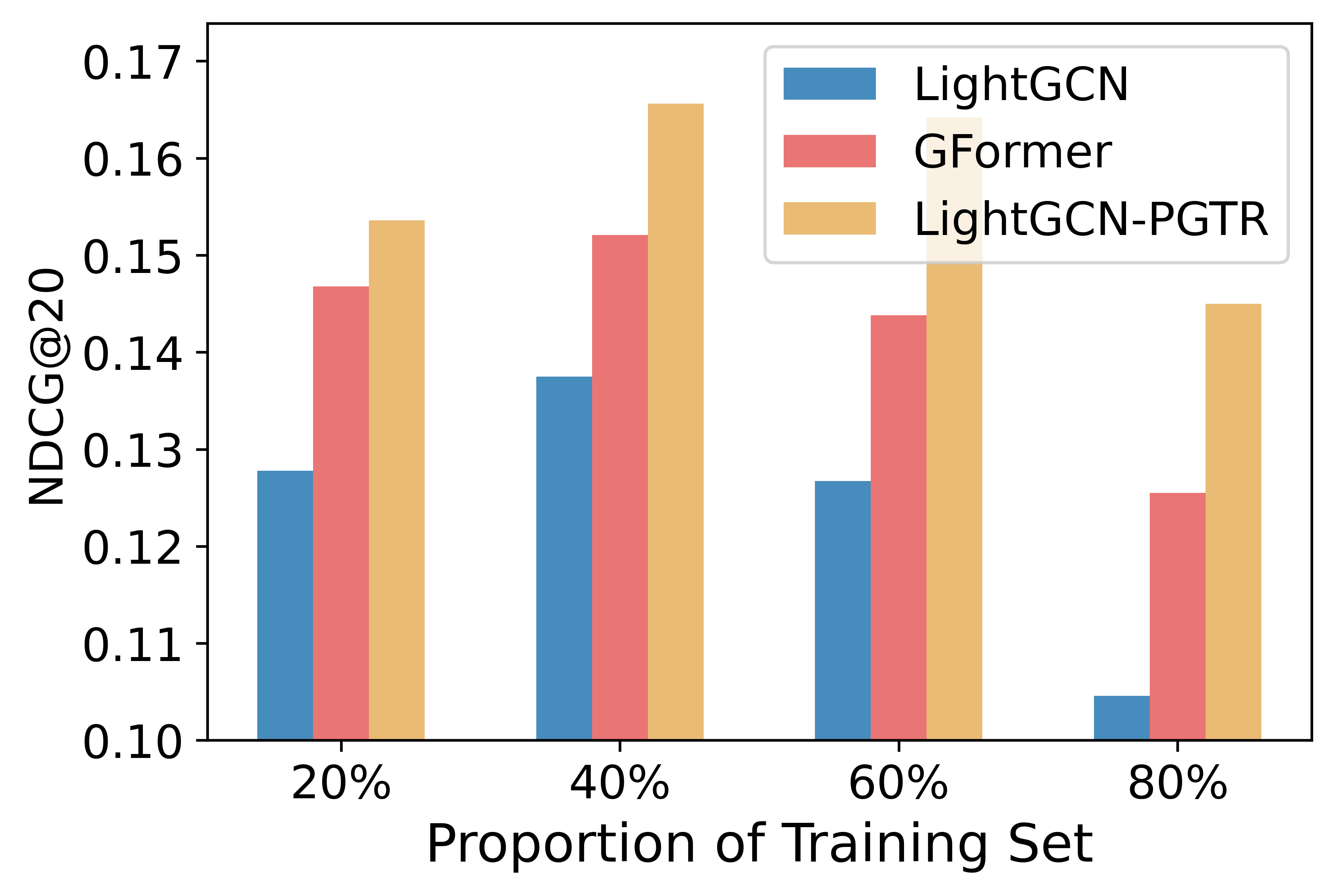}
    (b) Douban-book NDCG
\end{subfigure}%
\\
\begin{subfigure}[b]{0.45\textwidth}
    \centering
    \includegraphics[width=\textwidth]{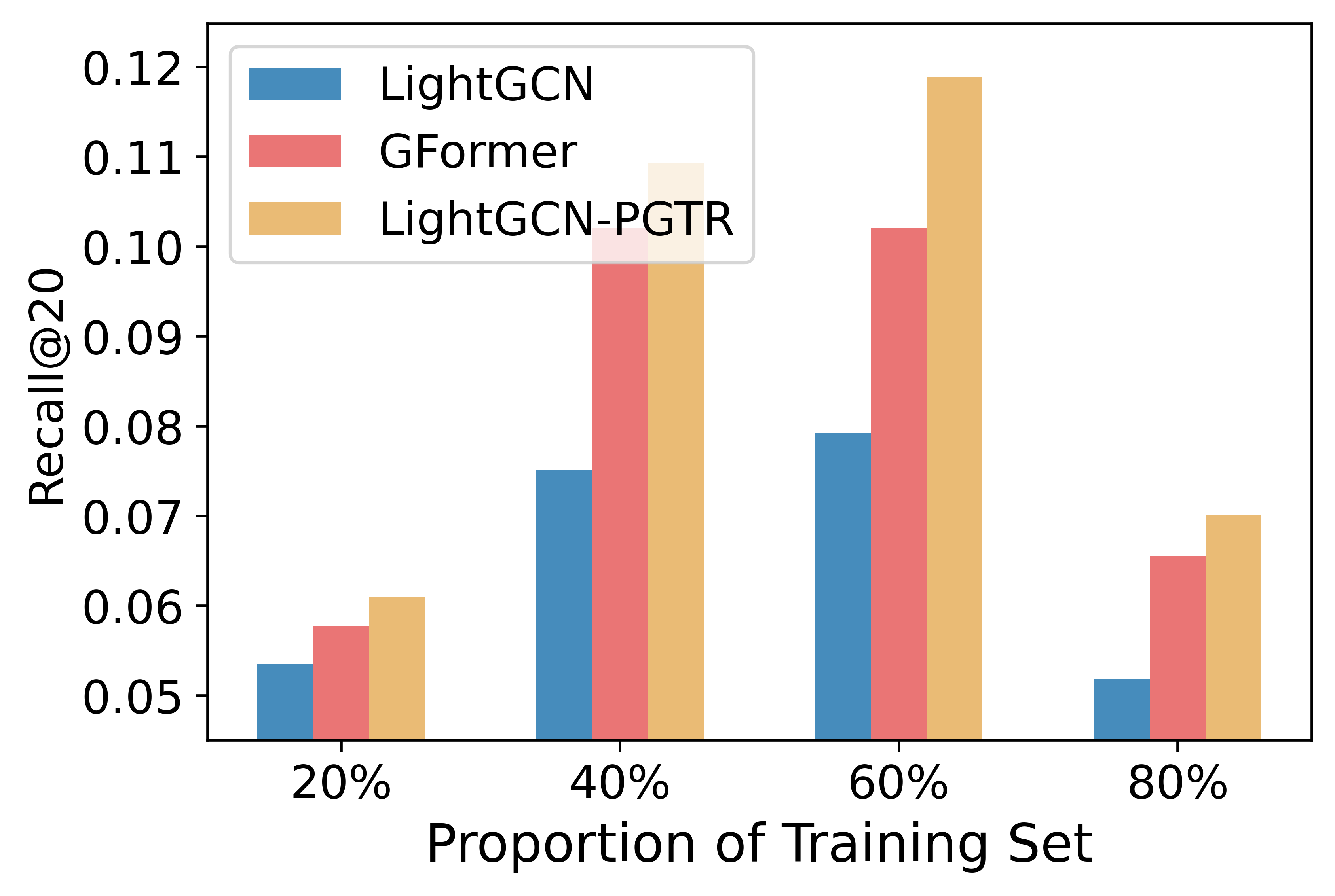}
    (c) LastFM Recall
\end{subfigure}%
\begin{subfigure}[b]{0.45\textwidth}
    \centering
        \includegraphics[width=\textwidth]{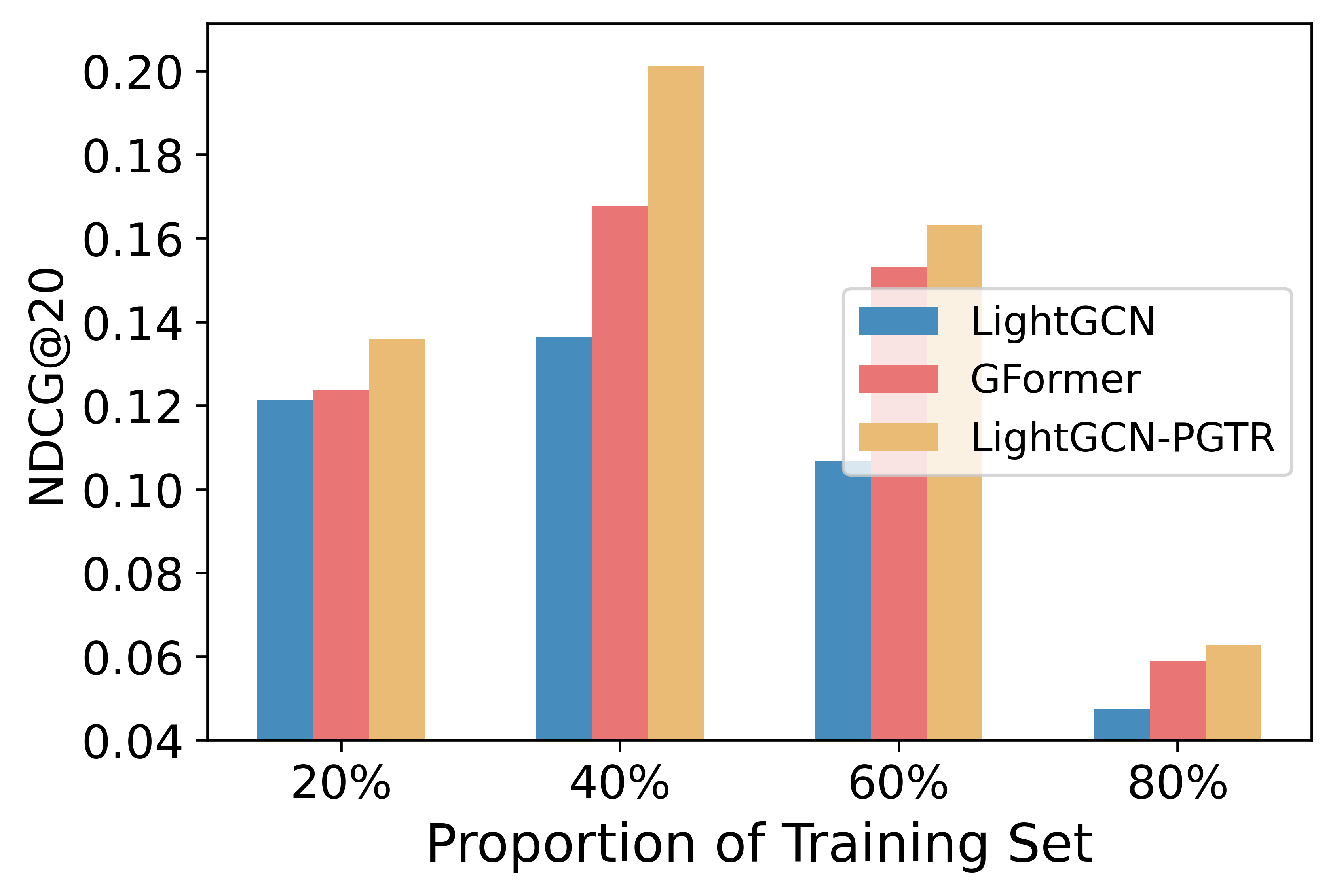}
    (d) LastFM NDCG
\end{subfigure}%
\caption{Performance comparison on Douban-book and LastFM datasets under different proportions of the training set.}  
\label{fig_sparsity}
\end{figure}

In order to evaluate the robustness of PGTR against data sparsity, we conduct various experiments on different sparsity levels. To be more specific, we conduct experiments on different proportions of training sets (20\%, 40\%, 60\%, 80\% of the original dataset, with the remainder used for testing), simulating various levels of sparsity. The results on two datasets, Douban-book and LastFm, are shown in Figure \ref{fig_sparsity}. From the figure, it can be observed that our method outperforms LightGCN and GFormer across different levels of sparsity, reaching the SOTA results. We believe that, in the current scenario of sparse recommendation data, adopting PGTR to model relationships among nodes explicitly and capture global collaborative relations is necessary.

\subsubsection{Robustness Against Data Noise} 
\begin{figure}  
\centering
\begin{subfigure}[b]{0.45\textwidth}
    \centering    
    \includegraphics[width=\textwidth]{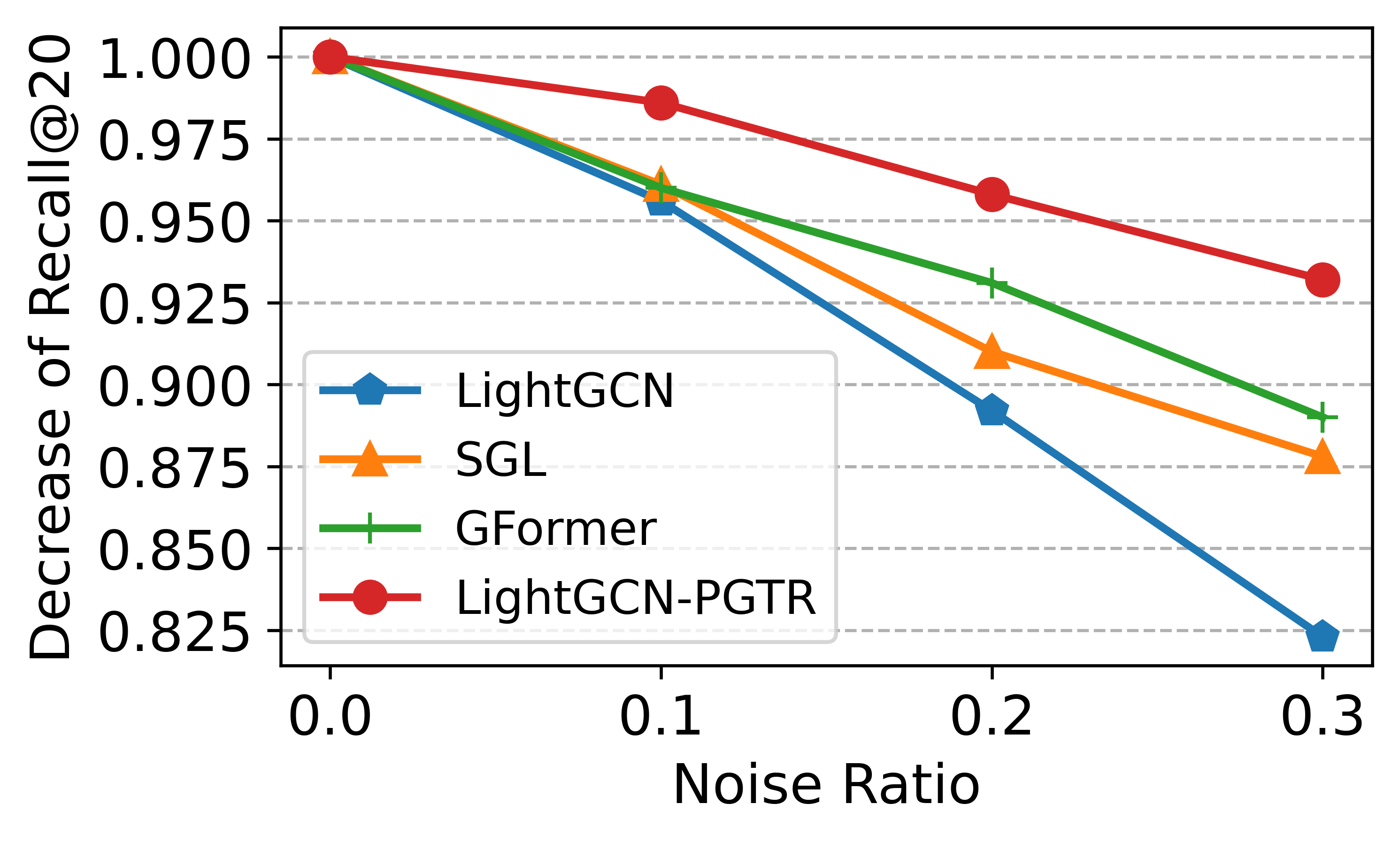}
    (a) Amazon-elec Recall
\end{subfigure}%
\begin{subfigure}[b]{0.45\textwidth}
    \centering
    \includegraphics[width=\textwidth]{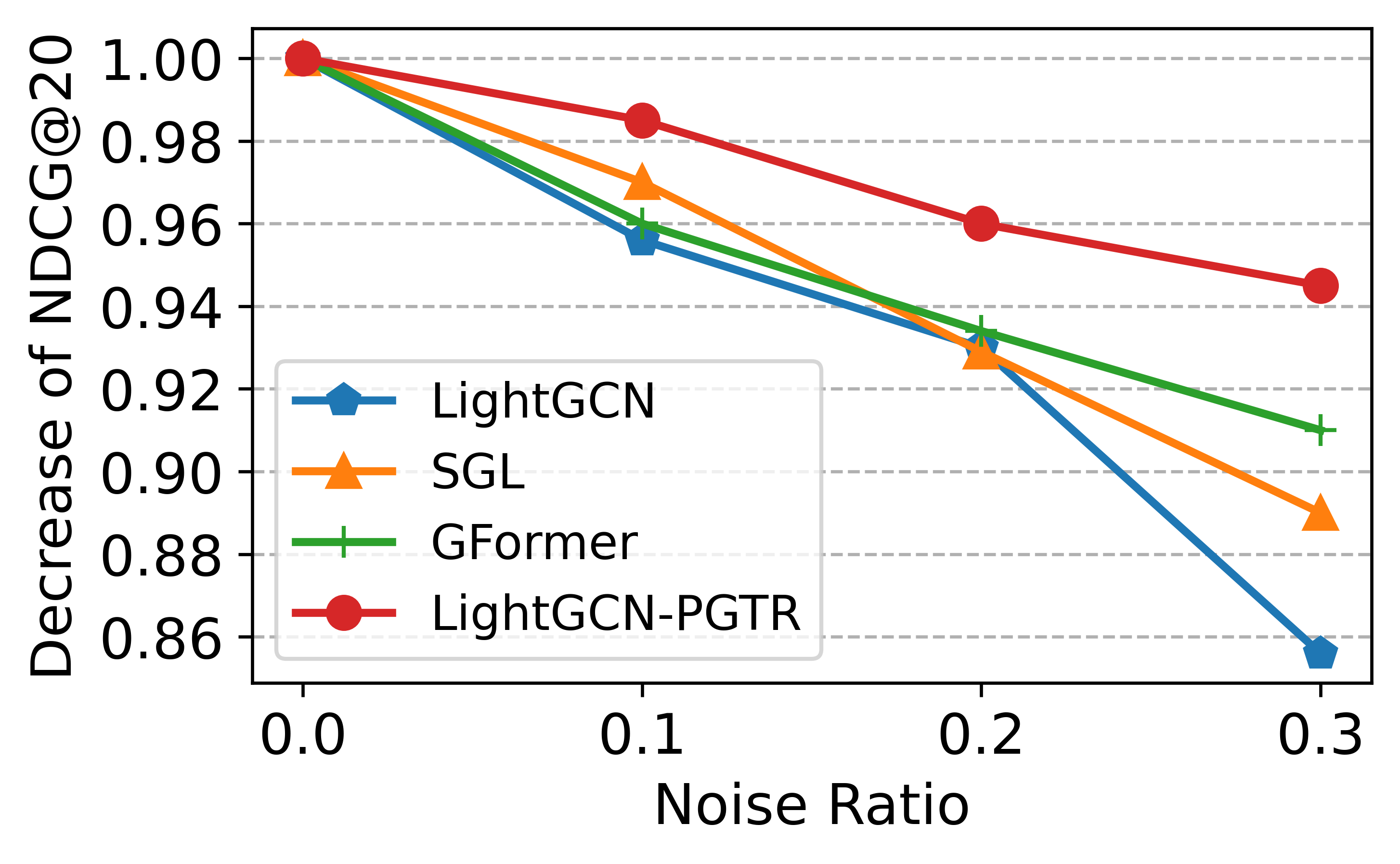}
    (b) Amazon-elec NDCG
\end{subfigure}%
\\
\begin{subfigure}[b]{0.45\textwidth}
    \centering
    \includegraphics[width=\textwidth]{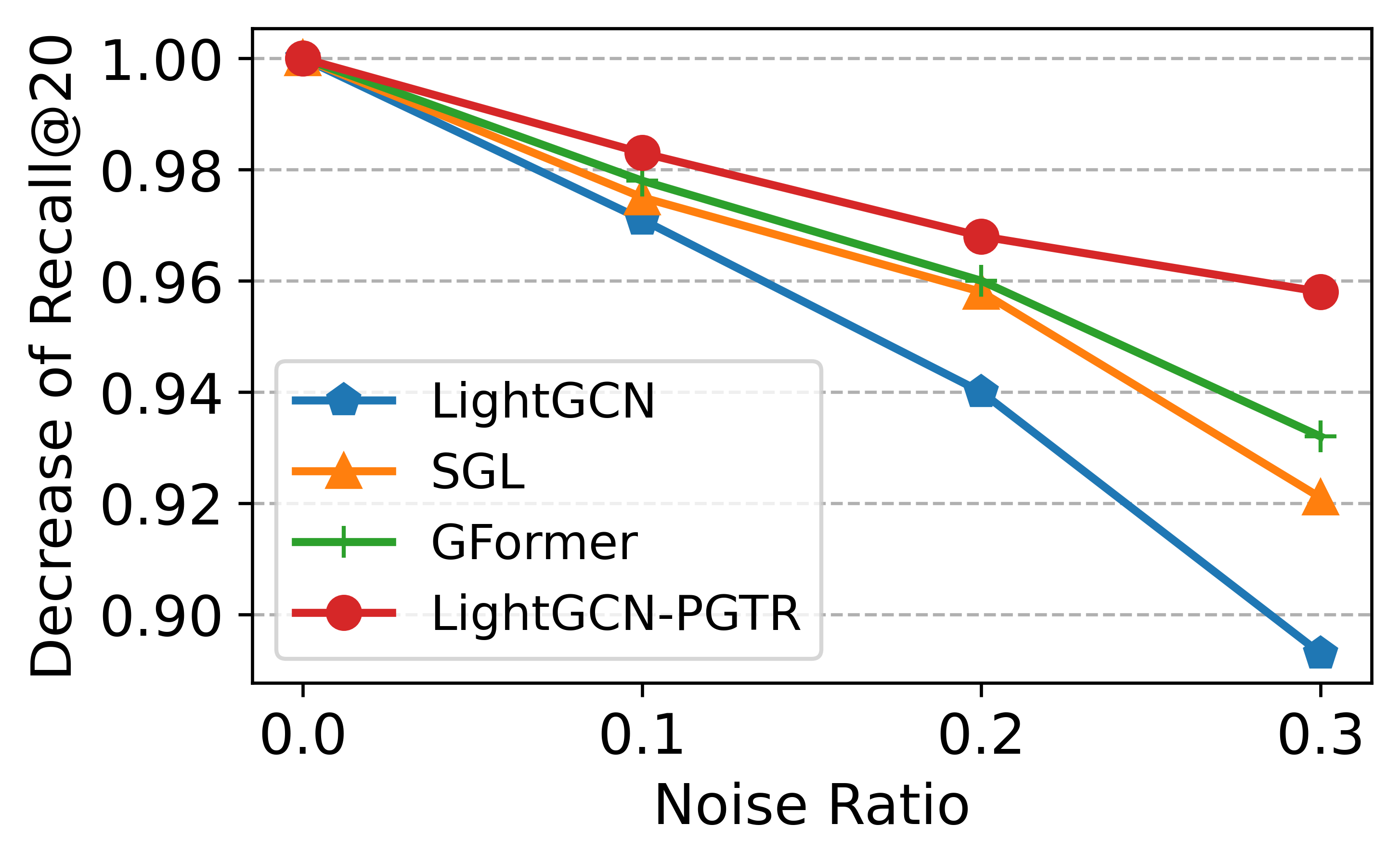}
    (c) Douban-book Recall
\end{subfigure}%
\begin{subfigure}[b]{0.45\textwidth}
    \centering
        \includegraphics[width=\textwidth]{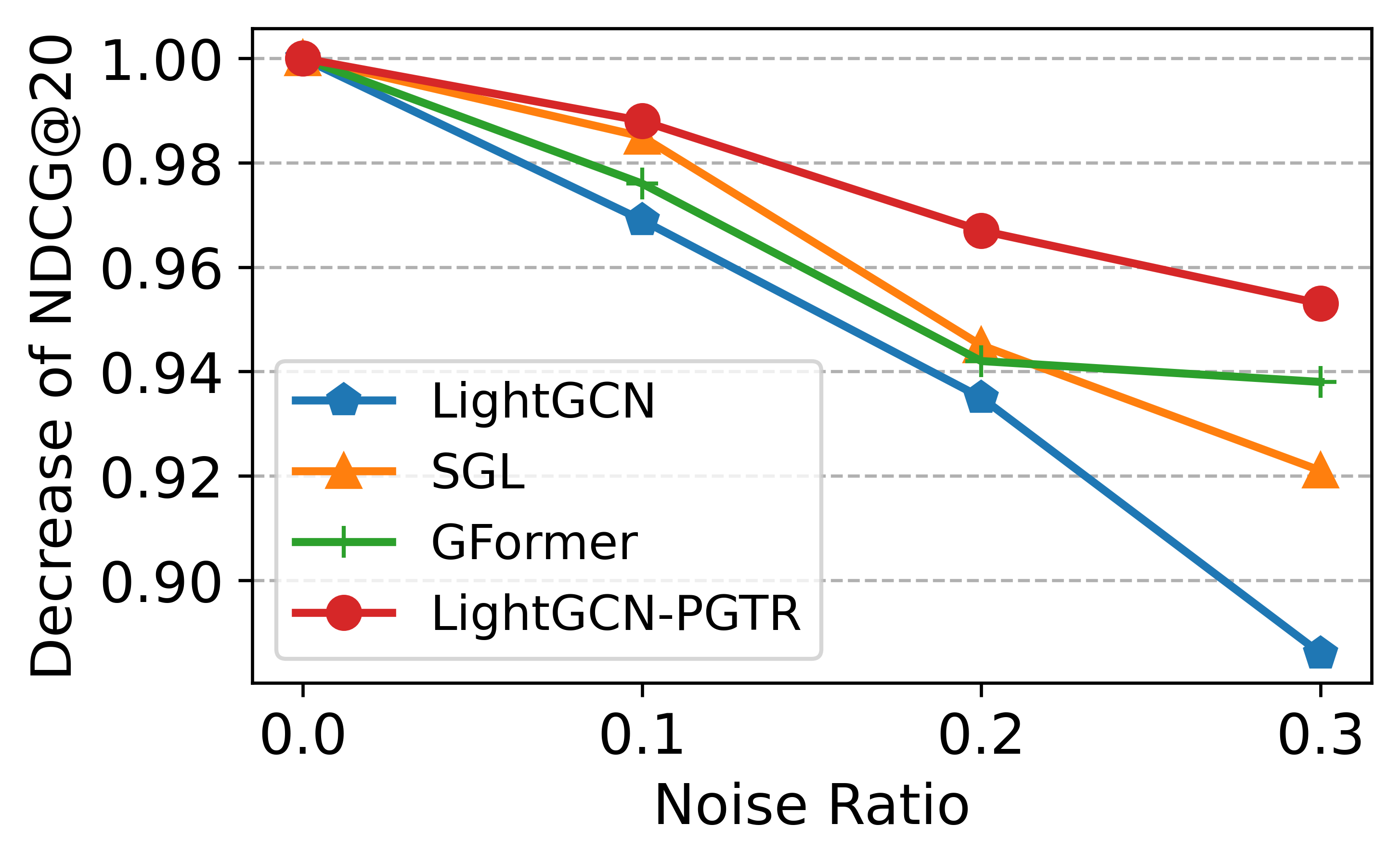}
    (d) Douban-book NDCG
\end{subfigure}%
\caption{Performance comparison on Amazon-elec and Douban-book datasets with different noise perturbations.}  
\label{fig_noise}
\end{figure}

In order to demonstrate the robustness of the proposed method PGTR against interaction noise, we randomly insert different proportions (10\%, 20\%, 30\%) of noisy interactions into the training set. The noisy interactions of a user are randomly generated based on a specific proportion to the number of interactions. We train PGTR and other powerful baselines and evaluate their performance. We compare the performance decrease of different methods under different proportions of noise. The performance without any noise is the benchmark, as shown in Figure \ref{fig_noise}. Firstly, we can observe that LightGCN had the highest decrease in performance on both datasets (Amazon-elec and Douban-book). This indicates that adopting this fixed neighbor aggregation is sensitive to noise. Secondly, it can be observed that SGL and GFormer have a certain ability to resist noise, which demonstrates the robustness of self-supervised learning against noise. Furthermore, the results show that our method has the least performance degradation on both datasets. This demonstrates that our method has good recognition capability for noise when dealing with interaction data. We believe that the PGTR model can integrate multiple positional information to discriminate the authenticity of interactions, which is the source of advancement in our method.

\subsection{Ablation Study (\textbf{RQ3})}
In this section, we conduct experiments to validate the effectiveness of the four proposed positional encodings and the role of capturing long-range/global collaborative information in our model.

\begin{figure*}
    \centering
    \includegraphics[width=0.85\textwidth]{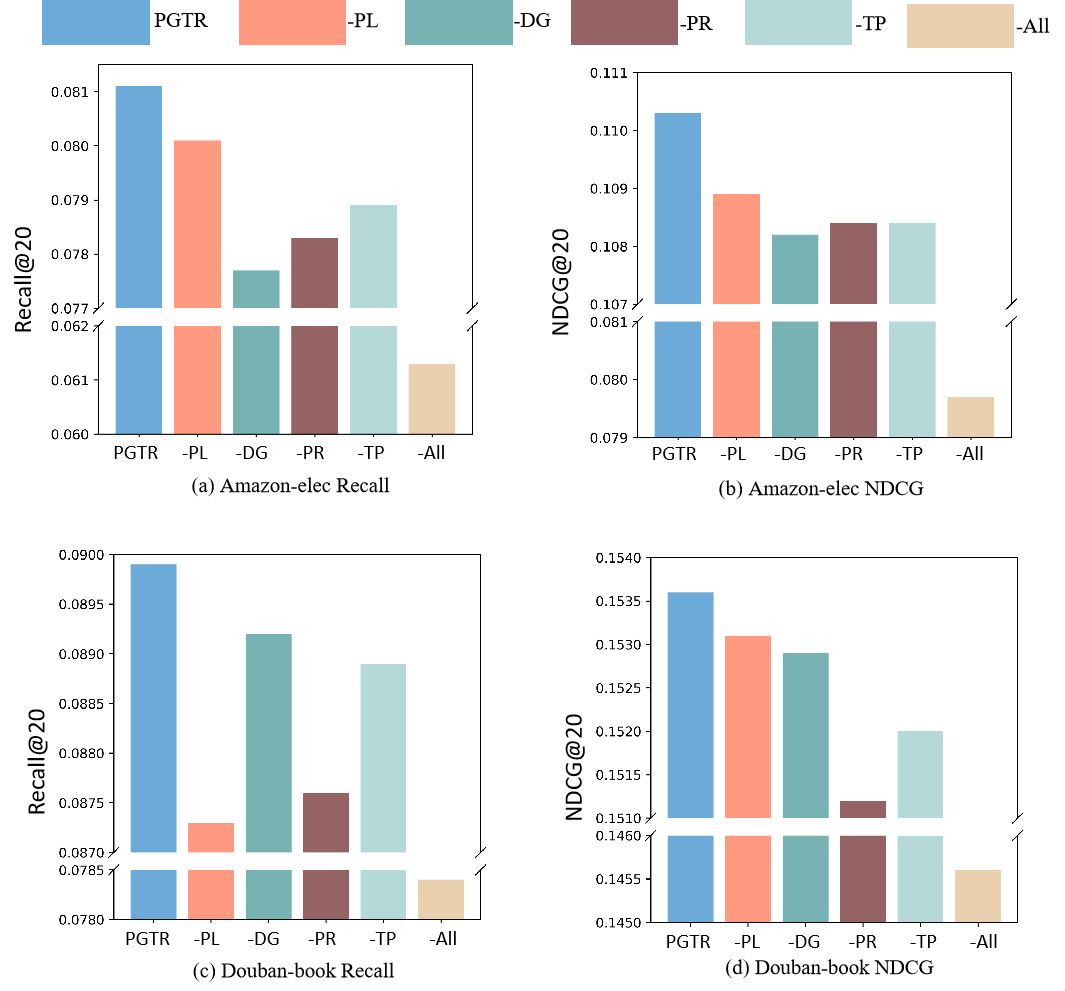}    
    \caption{Performance of ablated models on Amazon-elec and Douban-book datasets.}
    \label{fig_ablation_position}
\end{figure*}


\subsubsection{Effect of Positional Encoding} To validate the effectiveness of the proposed positional encodings, we perform an ablation study over them. Specifically, we separately remove each type of node positional encoding as well as all encodings from the PGTR and test the performance after their removal, i.e., '-PL', '-DG', '-PR', '-TP' and 'All' represent that removing spectral encodings, degree encodings, PageRank encodings, type encodings, and all encodings respectively. We show the results on Amazon-elec and Douban-book datasets in Figure \ref{fig_ablation_position}. Overall, it can be seen that after removing each positional encoding, the performance would degrade. It means that each positional encoding contributes to the PGTR model. In addition, each positional encoding can have varying effects on different datasets. For example, the impact of spectral encoding on performance is evidently greater in Douban-book compared to its effect on Amazon-elec. We can also observe that the significance of each positional encoding varies. This suggests that they play distinct roles in constructing relationships among nodes across different aspects. As a comparison, the performance drops significantly after all positional encodings are removed, which indicates that the proposed node positional encodings play an important role in the model.

\begin{figure}  
\centering
\begin{subfigure}[b]{0.5\textwidth}
    \centering    
    \includegraphics[width=\textwidth]{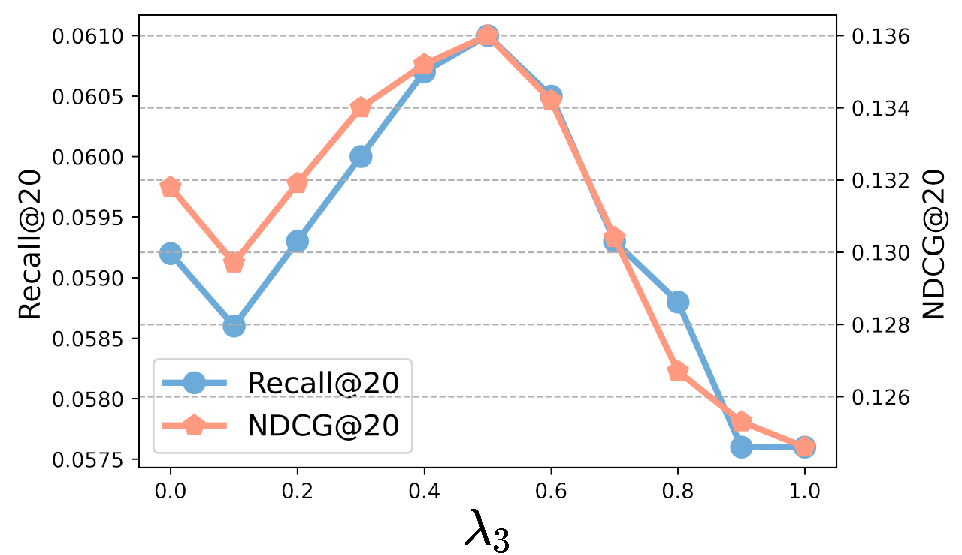}
    (a) LastFM
\end{subfigure}%
\begin{subfigure}[b]{0.5\textwidth}
    \centering
    \includegraphics[width=\textwidth]{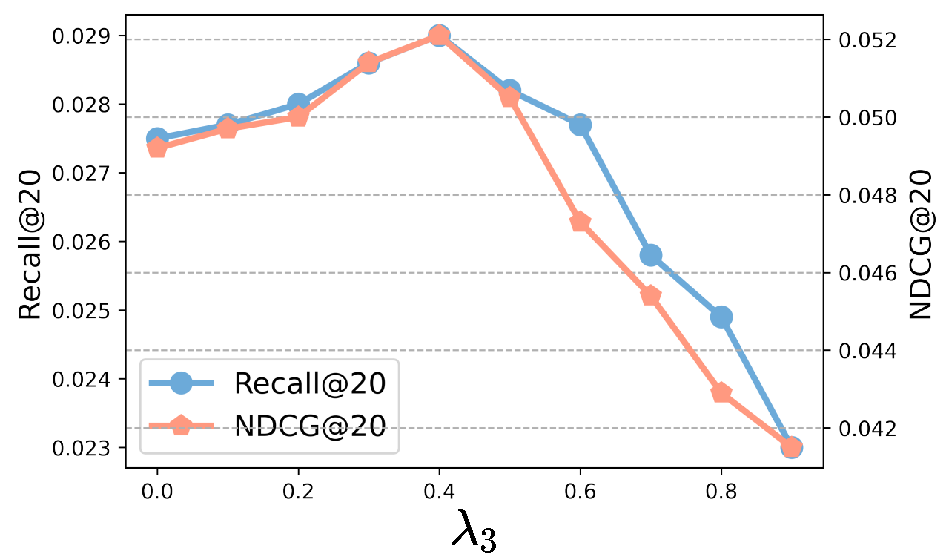}
    (b) Amazon-book
\end{subfigure}%
\caption{Effect of $\lambda_3$ on LastFM and Amazon-book datasets.}  
\label{fig_lambda3}
\end{figure}

\subsubsection{Effect of $\lambda_3$} $\lambda_3$ is used in Eq.\eqref{eq238101} to control the ratio of mixing between local and long-range/global collaborative information. We give the tuning results of $\lambda_3$ on LastFM and Amazon-book datasets in Figure \ref{fig_lambda3}. It demonstrates that global information plays a crucial role in performance. When $\lambda_3$ is small, the supplementation of local information is insufficient. When $\lambda_3$ is large, introducing excessive global information tends to introduce noise information and thus disrupt node representation learning. These situations can lead to a decrease in performance.

\subsubsection{Effect of the SSM loss} The SSM loss \cite{wu2022effectiveness} achieves a good performance in recommendation. To further validate the effectiveness of our framework, we compare the performance of the GCN-based model trained with SSM loss on the four datasets, as shown in Table \ref{tab_ssm}. It can be observed that SSM loss significantly improves the performance of the GCN-based model, demonstrating the importance of the self-supervised learning paradigm in recommendation. Furthermore, our approach still outperforms SSM significantly. This indicates that our method more effectively captures users' interests by modeling the relationships among users and items. We do not use the SSM loss on UltraGCN, as UltraGCN has been designed with a specialized loss function tailored to its unique architecture and requirements, thus there is no comparison with it. It should be noted that in both the '-SSM' and '-PGTR' methods, the temperature coefficient has been searched with the same granularity. Ultimately, the same setting has been applied to ensure a fair and comparable evaluation.


\renewcommand{\arraystretch}{1.5}
\begin{table*}[]
\centering
\caption{Effect of the SSM loss on the four datasets. Recall and NDCG scores are reported at a cutoff of 20.}
\begin{tabular}{c|cc|cc|cc|cc}
\hline
Data          & \multicolumn{2}{c|}{Amazon-elec} & \multicolumn{2}{c|}{Douban-book} & \multicolumn{2}{c|}{LastFM} & \multicolumn{2}{c}{Amazon-book} \\ \hline
Model         & Recall        & NDCG       & Recall        & NDCG       & Recall     & NDCG     & Recall       & NDCG       \\ \hline
NGCF          & 0.0684           & 0.0921        & 0.0735           & 0.1164        & 0.0263        & 0.0654      & 0.0141          & 0.0286        \\
NGCF-SSM      & 0.0758           & 0.1035        & 0.0735           & 0.1357        & 0.0587        & 0.1311      & 0.0260          & 0.0469        \\
\rowcolor[HTML]{EFEFEF} 
NGCF-PGTR     & 0.0802           & 0.1084        & 0.0909           & 0.1558        & 0.0616        & 0.1394      & 0.0274          & 0.0506        \\ \hline
GCCF          & 0.0358           & 0.0525        & 0.0570           & 0.1114        & 0.0325        & 0.0840      & 0.0179          & 0.0364        \\
GCCF-SSM      & 0.0732           & 0.1030        & 0.0844           & 0.1557        & 0.0447        & 0.0998      & 0.0254          & 0.0457        \\
\rowcolor[HTML]{EFEFEF} 
GCCF-PGTR     & 0.0789           & 0.1075        & 0.0889           & 0.1584        & 0.0579        & 0.1258      & 0.0264          & 0.0476        \\ \hline
LightGCN      & 0.0330           & 0.0485        & 0.0677           & 0.1278        & 0.0535        & 0.1215      & 0.0250          & 0.0472        \\
LightGCN-SSM  & 0.0613           & 0.0862        & 0.0788           & 0.1465        & 0.0582        & 0.1293      & 0.0273          & 0.0489        \\
\rowcolor[HTML]{EFEFEF} 
LightGCN-PGTR & 0.0811           & 0.1103        & 0.0899           & 0.1536        & 0.0610        & 0.1360      & 0.0290          & 0.0521        \\ \hline
\end{tabular}
\label{tab_ssm}
\end{table*}

\subsection{Computational Efficiency: GPU Utilization and Algorithm Runtime} 
\label{sec_compeff}
In the analysis of computational efficiency, as depicted in Table \ref{compeff_lastfm} and Table \ref{compeff_abook}, the proposed method PGTR demonstrates a trade-off between computational resources and performance. In this experiment, all methods are trained on an NVIDIA A40 (48GB) GPU and an Intel(R) Xeon(R) Gold 6326 CPU @ 2.90GHz. The training process is terminated if the performance did not improve for 20 consecutive epochs, and the total training time is recorded.
When compared to the baseline model LightGCN, LightGCN-PGTR exhibits a moderate increase in GPU utilization, with 2,552 MB and 2,972 MB on the LastFM and Amazon-book datasets, respectively. This is expected due to the additional positional encoding modules and the Transformer module's complexity. However, the single-round training time for LightGCN-PGTR (NGCF-PGTR) is approximately 5x that of LightGCN (NGCF-PGTR), which is attributed to the more complex model architecture and the computational overhead of capturing long-range dependencies. Despite this, the overall training duration of Light-PGTR is only 1.2x and 2.9x longer on the LastFM and Amazon-book datasets, respectively, when compared to LightGCN. Specifically, NGCF-PGTR even has a shorter overall training time than NGCF on the LastFM dataset. This indicates that while PGTR requires more computational resources and time per epoch, it converges faster, requiring fewer epochs to achieve greater performance. This efficiency in terms of epochs is crucial for large-scale recommendation tasks where the volume of data is extensive, and the model must generalize well from limited training data. Overall, the increased resource usage of PGTR is justified by its improved ability to capture long-range collaborative signals, leading to enhanced recommendation accuracy and robustness against sparsity and noise. 

\renewcommand{\arraystretch}{1.5}
\begin{table*}[]
\centering
\caption{Computational Efficiency of PGTR on the LastFM dataset.}    
\begin{tabular}{c|c|c|
>{\columncolor[HTML]{ECF4FF}}c |
>{\columncolor[HTML]{ECF4FF}}c }
\hline
LastFM                    & NGCF     & LightGCN & NGCF-PGTR & LightGCN-PGTR \\ \hline
GPU Utilization                  & 2,398 MB & 2,238 MB & 2,692 MB  & 2,552 MB      \\ \hline
Training Cost (per epoch) & 11.5s    & 11s      & 44s       & 44s           \\ \hline
Training Epochs           & 262      & 256      & 48        & 76            \\ \hline
Overall Training Cost     & 50.2min  & 46.9min  & 35.2min   & 55.7min       \\ \hline
Recall@20                 & 0.0263   & 0.0535   & 0.0616    & 0.0610        \\ \hline
\end{tabular}
\label{compeff_lastfm}
\end{table*}

\begin{table*}[]
\centering
\caption{Computational Efficiency of PGTR on the Amazon-book dataset.}   
\begin{tabular}{c|c|c|
>{\columncolor[HTML]{ECF4FF}}c |
>{\columncolor[HTML]{ECF4FF}}c }
\hline
Amazon-book               & NGCF     & LightGCN & NGCF-PGTR & LightGCN-PGTR \\ \hline
GPU Utilization           & 2,676 MB & 2,376 MB & 3,188 MB  & 2,972 MB      \\ \hline
Training Cost (per epoch) & 13s      & 11.5s    & 59 s      & 67s           \\ \hline
Training Epochs           & 118      & 254      & 81        & 128           \\ \hline
Overall Training Cost     & 25.6min  & 48.7min  & 79.7min   & 142.9min      \\ \hline
Recall@20                 & 0.0141   & 0.0250   & 0.0274    & 0.0290        \\ \hline
\end{tabular}
\label{compeff_abook}
\end{table*}

\section{Related work}
In this section, we review related works that include graph-based recommendation and positional encodings of graph transformer.

\subsection{Graph-based Recommendation}
Graph-based recommendation approaches have gained attention for effectively capturing complex relationships among users and items, making them suitable for personalized recommendations\cite{wu2022graph,chen2022gdsrec,chen2023graph}. These methods leverage the inherent structure and connectivity of data to enhance recommendation performance. Various algorithms like NGCF \cite{wang2019neural}, LightGCN \cite{he2020lightgcn}, UltraGCN \cite{KelongMao2021UltraGCNUS}, etc, exploit the graph's information for recommendation tasks. NGCF \cite{wang2019neural} utilizes graph convolutional networks to capture the complex patterns in user-item interactions. It propagates user and item embeddings through the graph to refine their representations, which are then used for making recommendations. LightGCN \cite{he2020lightgcn} is a simplified version of GCN designed specifically for recommendation tasks. It removes complex operations such as feature transformation and nonlinear activation, focusing solely on the essential component of neighborhood aggregation. This results in a model that is not only easier to train but also achieves significant improvements in recommendation performance over more complex models like NGCF. UltraGCN \cite{KelongMao2021UltraGCNUS} is another advancement in graph-based recommendations. It introduces an ultra-high-order proximity preserving scheme that captures long-range dependencies in the user-item graph. This allows the model to better understand the intricate connections between users and items, leading to enhanced recommendation accuracy. However, these methods are constrained by local information and lack direct and explicit modeling of long-range collaborative signals, which can lead to performance loss.
Recently, there have been several studies aiming to encode global information into local representations based on contrastive learning~\cite{yang2021enhanced,cai2022lightgcl,xia2022hypergraph}. By this way, nodes can obtain long-range collaborative signals. For example, EGLN~\cite{yang2021enhanced} proposes a local-global consistency optimization function to capture the global properties learned from the augmented interaction graph, which is obtained by calculating the similarities between users and items, retaining several largest similarities~\cite{yu2023self}. This approach allows the model to better understand the overall structure of the user-item interactions, leading to improved recommendations even for users and items with sparse interaction histories.
LightGCL~\cite{cai2022lightgcl} utilizes the principle components of the interaction graph to distill important collaborative signals from the global perspective and then refines node representations by local-global contrastive learning. This method is particularly effective in scenarios where the interaction graph is dense and high-dimensional, as it can capture the underlying patterns that are not immediately apparent from local connections alone.
However, these methods heavily rely on the way of constructing global collaborative relations and may lose some important signals. The construction of the augmented interaction graph in EGLN~\cite{yang2021enhanced}, for instance, depends on the selection of top similarities, which can be sensitive to the choice of threshold and may exclude some valuable interactions. Similarly, LightGCL~\cite{cai2022lightgcl} depends on the accurate identification of principal components, which may not always capture the most relevant aspects of the global structure, especially in cases where the interaction graph is noisy or incomplete.

Additionally, auxiliary information, such as item attributes, user demographics and knowledge graph (KG) is incorporated into the graph representation to capture diverse preferences and improve recommendation accuracy \cite{li2019fi,cao2019unifying,wang2019knowledge}. Recently, self-supervised graph-based methods have been widely explored and utilized in the field of recommendation systems \cite{wu2021self,yu2022graph,xia2022hypergraph,lin2022improving,yang2022knowledge}. They facilitate the discovery of meaningful structures and embeddings within the graph, leading to improved recommendations.

The GT has gained popularity in multimedia recommendation \cite{wei2023lightgt}, multi-behavior recommendation \cite{xia2021knowledge}, and sequential recommendation \cite{fan2021continuous}. However, its exploration in the field of collaborative filtering remains limited. GFormer \cite{DBLP:conf/sigir/LiXRY0023} and SHT \cite{xia2022self} utilize the GT for global relations modeling. However, the modeling is not adequate. In this work, we explore using the GT to capture collaborative relationships between all pairs of nodes, which is different from them.

\subsection{Positional Encoding of Graph Transformer}

Positional encoding (PE) is a critical technique in graph transformers (GT) that embeds positional information into the model, allowing it to understand the relative positions of nodes within a graph. This is essential because the transformer architecture, which GTs are based on, does not inherently account for order or position. Various forms of positional encoding have been explored to enhance the performance of GTs in recommendation systems.

Laplacian positional encoding is one such method that uses the graph's Laplacian matrix to capture the structural properties of the graph \cite{dwivedi2020generalization, kreuzer2021rethinking}. This encoding helps in understanding the role of each node within the graph's structure, providing a sense of the node's "position" in terms of its connectivity to other nodes. Shortest-path-distance encoding is another approach that considers the shortest path distances between nodes as a measure of their relative positions \cite{li2020distance}. This is particularly useful in recommendation systems where the goal is to recommend items that are "close" to the user's current preferences, even if they are not directly connected. Centrality-based encoding leverages the concept of node centrality, which measures the importance or influence of a node within the graph \cite{ying2021transformers}. Nodes with high centrality are likely to be more influential in shaping the recommendations, as they are more connected to other nodes. In the context of collaborative filtering, positional encoding can help in distinguishing between nodes that have similar interaction patterns but different positions within the graph. For example, two users may have similar preferences, but their positions in the social graph or their influence within the community might differ, affecting the recommendations they receive.

To the best of our knowledge, there has been limited work on utilizing positional encoding in collaborative filtering scenarios. GFormer \cite{DBLP:conf/sigir/LiXRY0023}, in particular, only uses the distance between the target node and sampled anchor nodes to preserve global topological structure information. We argue that this is insufficient. Exploring how to effectively encode positional information in collaborative filtering is a new question. In this paper, we design several simple yet effective positional encodings tailored for the user-item interaction graph in collaborative recommendation.

In summary, positional encoding in graph transformers for recommendation systems is an evolving field with promising potential. By providing explicit positional information, these encodings help models capture the intricate spatial dependencies within user-item graphs, leading to more personalized and accurate recommendations.

\section{Conclusion and future work}
In this paper, our focus is incorporating the Graph Transformer into collaborative recommendation to address an important problem existing in the GCN-based methods: capturing long-range collaborative signals. To this end, we propose the PGTR framework which can be applied to various GCN-based backbones. The key contribution of this study lies in the design of node positional encodings tailored for the interaction graph in collaborative recommendation within PGTR for learning relationships between nodes. We conduct experiments on multiple datasets, demonstrating the effectiveness of our approach.

In the future, we will design more effective positional encodings for different interaction graphs, exploring how to capture graph structural information and a variety of node relationships. Furthermore, we will explore the application of GT in various recommendation scenarios, addressing issues of data sparsity and node dependency in recommendations.


\bibliographystyle{ACM-Reference-Format}
\bibliography{sample-base,reference}


\end{document}